\documentclass[sigconf]{acmart}

\AtBeginDocument{%
  }

\acmISBN{978-1-4503-XXXX-X/18/06}
\copyrightyear{2024} 
\acmYear{2024} 
\setcopyright{rightsretained} 
\acmConference[CHI '24]{Proceedings of the CHI Conference on Human Factors in Computing Systems}{May 11--16, 2024}{Honolulu, HI, USA}
\acmBooktitle{Proceedings of the CHI Conference on Human Factors in Computing Systems (CHI '24), May 11--16, 2024, Honolulu, HI, USA}
\acmDOI{10.1145/3613904.3642646}
\acmISBN{979-8-4007-0330-0/24/05}

\graphicspath{{figures/}}
\usepackage{caption}
\usepackage{subcaption}
\usepackage{multirow}
\usepackage{todonotes}

\usepackage{algpseudocode}
\usepackage{algorithm}

\begin{document}

\title{Comparison of Spatial Visualization Techniques for Radiation in Augmented Reality}

\author{Fintan McGee}
\email{fintan.mcgee@list.lu}
\orcid{0000-0001-7398-2664}
\affiliation{%
  \institution{Luxembourg Institute of Science and Technology}
  \streetaddress{Maison d'Innovation, 5 Avenue Des Hauts-Fourneaux}
  \city{Esch-sur-Alzette}
  \country{Luxembourg}
  \postcode{L-4362}
}
\author{Roderick McCall}
\email{roderick.mccall@list.lu}
\orcid{0000-0002-0765-8919}
\affiliation{%
  \institution{Luxembourg Institute of Science and Technology}
  \streetaddress{Maison d'Innovation, 5 Avenue Des Hauts-Fourneaux}
  \city{Esch-sur-Alzette}
  \country{Luxembourg}
  \postcode{L-4362}
}
\author{Joan Baixauli}
\email{joan.baixauli@list.lu}
\orcid{0000-0001-7327-0966}
\affiliation{%
  \institution{Luxembourg Institute of Science and Technology}
  \streetaddress{Maison d'Innovation, 5 Avenue Des Hauts-Fourneaux}
  \city{Esch-sur-Alzette}
  \country{Luxembourg}
  \postcode{L-4362}
}

\renewcommand{\shortauthors}{McGee et al.}

\begin{abstract}
  Augmented Reality (AR) provides a safe and low-cost option for hazardous safety training that allows for the visualization of aspects that may be invisible, such as radiation. Effectively visually communicating such threats in the environment around the user is not straightforward. This work describes visually encoding radiation using the spatial awareness mesh of an AR Head Mounted Display.  We leverage the AR device’s GPUs to develop a real time solution that accumulates multiple dynamic sources and uses stencils to prevent an environment being over saturated with a visualization, as well as supporting the encoding of direction explicitly in the visualization. We perform a user study (25 participants) of different visualizations and obtain user feedback. Results show that there are complex interactions and while no visual representation was statistically superior or inferior, user opinions vary widely. We also discuss the evaluation approaches and provide recommendations.
\end{abstract}



\received{15 September 2023}
\received[revised]{12 December 2023}
\received[accepted]{22 February 2024}

\begin{teaserfigure}
    \includegraphics[trim={0 0 0 5cm},clip,width=\textwidth]{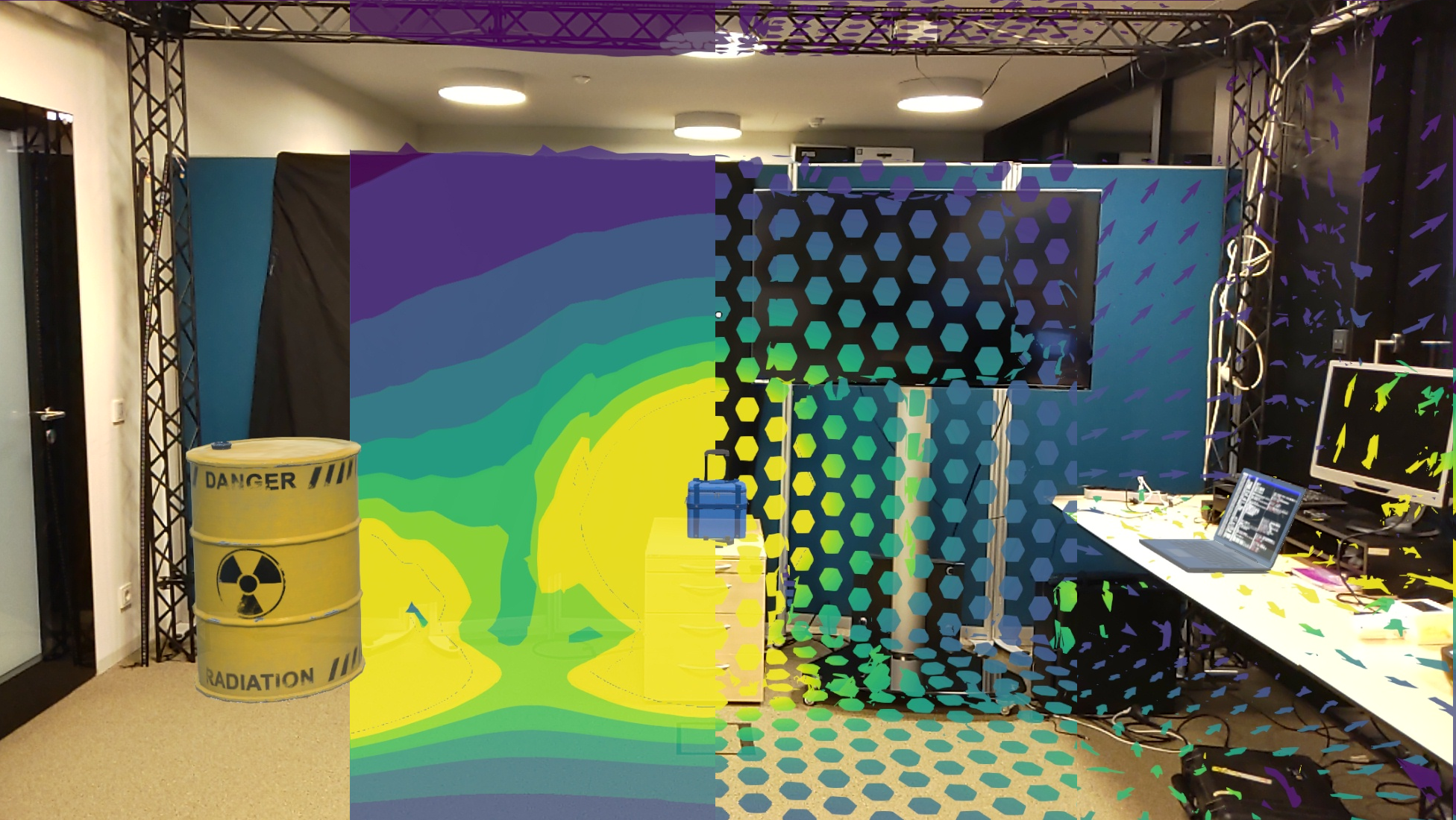}
    \caption{Three different approaches to visually encoding radiation in the physical environment around the user. In this composite image there are three Augmented Reality radiation sources, the barrel (left), the small suitcase (middle) and the laptop on the table (right).  At the leftmost side there is no visualization of the radiation, moving right, a banded visualization coloring the spatial awareness mesh is used, next a hexagon is used as a stencil, to allow better visibility of the real-world objects. Finally, an arrow-based visualization is shown illustrating the flow of radiation away from the sources.}
    \Description{ A composite image of a room showing three different augmented reality visualizations of radiation from virtual sources.}
    \label{fig:teaser}
  \end{teaserfigure}

\maketitle

  © Fintan McGee, Roderick McCall, and Joan Baixauli. 2024. This is the author's version of the work. It is posted here for your personal use. Not for redistribution. The definitive Version of Record was published in Proceedings of the CHI Conference on Human Factors in Computing Systems (CHI ’24), \url{https://doi.org/10.1145/3613904.3642646}.

\section{Introduction}
AR offers the possibility to show people make the invisible, visible.  It allows a user to see the inside of a printer to help with maintenance tasks \cite{Feiner1993}, or to visualize underground civil infrastructure \cite{Zollmann2010}, while archaeological AR can allow long disappeared buildings to be visible once again in their original setting \cite{Vlahakis2002}. AR can also enable a user to understand other aspects of their environment that may not be visible to the naked eye such as pollution \cite{White2009} or radiation~\cite{Leucht2015}. These aspects may be completely simulated, as part of a training exercise, but visualizing them helps a user better understand their environment and the results of their actions. AR has long been used for training, particularly in situations where training with real material and equipment may be prohibitively costly and possibly even dangerous, e.g., laboratory safety training. In this paper, we examine the use of AR for the visualization of environmental threats, specifically harmful radiation (see Figure \ref{fig:teaser}). Using simulated virtual radiation sources  avoids the risk of exposure of training participants to actual radiation.
While our approach has been motivated by the use of AR for the training of emergency responders for radiological incidents, we believe the techniques and results described here can be generalized for other applications.

It is possible to visualize radiation by highlighting the source object, but of course the radiation field extends beyond its physical boundaries. Visualizing a field is also difficult, not least because the extents can span large areas, and it is difficult to convey the increase of intensity at points closer to the source. Therefore, we choose to visualize the intensity of radiation on the surfaces around the sources and the user. To this end we leverage the spatial awareness mesh, provided by the HoloLens 2 AR Head Mounted Display (HMD). However, this also presents challenges. AR blends the virtual with the physical. A simple coloring of the mesh and objects around the user based on intensity of radiation may obscure objects in the real world.
The safety critical nature of the work means that the visualization should be easily understood and imply the correct (or safest) movement of the operative concerned within the space. Also given the life critical time constraints involved, the visualization should not require time to be understood, it should be almost implicit.
Our approach can be considered as situated  visualization, as the radiation is visualized around the position of a virtual source in the real world. 
In the context of the recent work of Lee \emph{et al.}~\cite{lee2023} on situated visualization, such an approach can be considered a \emph{decal} design pattern, in the sense that it is a texture mapping applied to the real-world physical objects modeled by the spatial awareness mesh. 
\subsection*{Objectives and Motivation}
The objective of this work is to demonstrate how radiation can be best visualized in a dynamic real-time manner in the environment around the user. The purpose of visualization is to help a human user to better understand the nature of the threat they are facing, allowing them to minimize their radiation exposure. 
The scenario that motivates our work is  the training of first responders for radiological incidents. Where the purpose is to find, identify and retrieve the radiological sources. The purpose of the visualization in this training scenario is to provide team members with an understanding of the radiation distribution in the room from multiple virtual radiation sources.
The basic reasoning being that the movement of the user, and hence their proximity to radiation sources and the time spent near the sources, will vary based on the visualization approach. The most effective visualization being the one where the total dose of radiation absorbed is the lowest. We explore techniques to give a user the best perception of both the virtual radiation and the real word environment around them (see Figure \ref{fig:teaser}).
 The solution must work in real time as a radioactive source may be moved around the scene. 
Additionally, there may be multiple sources so the effects of the radiation must be visualized in a cumulative manner.  We want to avoid obfuscating an entire scene and physical environment while also ensuring that the intensity of the sources is comprehensible to the user. The visualization should help the user understand which potential path will result in the lowest level of exposure.
The contribution of this work lies in the technical demonstration of the approach implemented on commodity hardware, the empirical evaluation comparing the resulting visualizations in a user study, as well as the discussion concerning the results and evaluation approach.

\section{Background and related work}
In this section we describe related visualization work in AR, focusing on CBRN (Chemical, Biological, Radiological and Nuclear) response and radiation visualization.
\subsection{Augmented Reality and Visualization}
Spatial integration of virtual objects into the real world, in the context of visualization in AR, has been recognized as a challenge for some time \cite{Kalkofen2011}.  It can be done as part of an in-situ visualization, where the data and visualization are co-located in the real world, such as pollution data in a city \cite{White2009}.  In-situ visualization AR also is used to visualize data that is inherently 3D such as spatio-temporal interaction data, as done by the MIRIA Toolkit \cite{Buschel2021}.  Recently the topic has received renewed attention in the literature, such as the work of Lee \emph{et al.} \cite{lee2023} describing design patterns for situated visualizations, and the work of Calepso \emph{et al.} \cite{Calepso2023} exploring AR for situated analytics. Other work has also shown that 3D Data manipulation in AR with tangible markers improves time and accuracy for tasks that require coordination between perception and interaction \cite{Bach2018}. The motivating factor for this work is to provide a safe training environment (e.g., no active radiation sources), where the visualization techniques can improve future safe behaviors of the end-users.
\subsection{AR for CBRN Response}
AR and VR have been used to support training for a wide range of potentially dangerous incidents e.g., US Army \cite{ARES2018} and CBRN device training using VR \cite{ForgeFx2023}.   AR has also been used for managing actual radioactive incidents, for example  \cite{Ismael2023} to provides live information on the source, position and spread of radioactive materials.  The primary advantage of using AR is that it makes the invisible, e.g., radiation, visible. Furthermore, live information from simulated or real equipment can be provided in real-time to the HMD.

Typically, radioactive incident teams must (1) find sources (2) identify the type of source (3) extract the source from the scene. The procedures vary in each country, and for safety reasons the extraction team is not the same one that is used to find and locate the sources. In our scenario, we assume that the sources have been found, identified and that the operative assumes the role of an extraction team member who should remove the source  from the scene. In order to fulfill this task, the operative must have a good level of  situation awareness \cite{Endsley1988}, which relies on three key aspects (a) perception (b) comprehension (c) planning. If they possess a good level of situation awareness, they should be able to plan the best path to the source and remove it from the scene while minimizing the total amount of radiation absorbed. The focus in our visualization approaches is predominantly on egocentric perception. By this we mean how the operative estimates the distance between themselves and the stimuli (radiation source). 
\subsection{Existing AR Approaches for Visualizing Radiation}
Guarese \emph{et al.} \cite{Guarese2021} visualize electromagnetic fields in AR for the purpose of compatibility testing. Their approach uses lines in 3D space to show a 3D field and colored vectors to show emitted electromagnetic radiation.
Electromagnetic signals are also the focus of WaveRider~\cite{Rowden2022}. It visualizes signals of WiFi routers, using the surfaces (walls, ceilings, and floors) of the environment around the user. The authors use novel visualization approaches and have developed and evaluated their techniques in VR using a 3D model of the environment. They have also developed an initial prototype in AR, although for the AR prototype the 3D model for the visualization is manually registered with the real world. In contrast in our work, it is automatically registered.
Recent work by Meireles \emph{et al.} \cite{Meireles2023}, which extended the prior work of Carmo \emph{et al.} \cite{Carmo2017}, focused on situated visualization of solar radiation on building facades using tablet based AR. Building data is retrieved from a spatial database, based on the user position and search radius, and parallelepiped glyphs are drawn over the facades, with an artificial color scale to indicate the underlying values.
VIPER \cite{smith2022} visualizes the fields of static radiation sources as semi-transparent isocontours using a rainbow based colormap to show different intensities of radiation. However, it is difficult to understand the radiation levels at different points in space, therefore the isocontour is sliced into on the vertical and horizontal axis. Additionally, the isocontours may also obscure other aspects of the scene.  VIPER also visualizes a user's path to see the radiation intensity at each point.
Leucht \emph{et al.}~\cite{Leucht2015} describe an approach to visualizing X-ray radiation in a surgical context. Their work focuses on simulating radiation doses from  a C-Arm fluoroscope used by surgeons and uses two depth cameras to create a visual representation of the surgeon augmented with a color coded radiation map. The spatial mesh is generated from a stationary depth camera, and the setup process requires  calibration. There is no HMD, and the fluoroscope is the only source of radiation, and it cannot move around the scene.
Most recently Pakari \emph{et al.}~\cite{Pakari2023} delivered a solution for real time visualization of radiation data output from a dual particle radiation detector. They use a perceptually uniform colormap rendered on the spatial awareness mesh around the detector, as well as using a ray-based approach. Their approach is targeted towards real radiation data being processed and visualizing the reported radiation levels in the space around the detector. It is not targeted towards training purposes with multiple virtual sources.

\section{Technical Approach}
In this section we describe the requirements for our visualization approaches, and the techniques we used to realise them.
\subsection{Requirements}
Our technical approach is constrained by the fact that it is integrated as part of a larger existing AR training platform. This platform uses the HoloLens application as a client to a central server application and provides different training scenarios. We did not have the option of using a toolkit such as RagRug~\cite{Fleck2023} or DXR~\cite{Sicat2019}, as our approach needed to not require significant changes to the existing architecture and code base and be flexible enough to seamlessly integrate into other scenarios. Furthermore, trainees need to interact in the real spaces, unobstructed by AR visualizations. We cannot use a volumetric visualization approach or 3D isocontour approach as done by VIPER~\cite{smith2022}, as this will clutter to the scene and obscure real-world items. Our approach of drawing on the physical environment to reduce clutter  has some similarities to the approach of WaveRider~\cite{Rowden2022}, although the environment in that case is defined by an existing 3D model and our use cases are quite different.
To meet our objectives for supporting the training for radiological incidents, the following requirements, based on first responder feedback during a previous project, had to be met by the implementation:
\begin{enumerate}
\item The visualization should help the user determine the safest path to follow. 
\item The radiation will be visualized on the surfaces around the sources (i.e., using the spatial awareness mesh), to avoid adding extra clutter to the scene.
\item  The visualization should be responsive and support real-time updates of moving sources.
\item  Radiation must accumulate from multiple sources.
\item It must be possible to allow the users to see both the physical world and the virtual radiation.
\end{enumerate}

The key to visualizing the spatial awareness mesh using a texture that reflects the radiation intensity at that given point in three dimensions.  There are many issues to be overcome such as: (1) accessing the mesh data points, (2) the lack of texture coordinates for the continuously updated spatial awareness mesh, (3) the mapping between a texture point on the 2D mesh, (4) the appropriate level of radiation intensity and, (5) orienting a texture towards a 3D point, if orientation is being encoded.

\subsection{Hardware and Using the Spatial Awareness Mesh for Visualization}
The HoloLens 2 was chosen as it is one of the more advanced headsets on the market and offers the spatial awareness functionality required by our approach. It is an HMD that allows the users nearly full visibility of the real world, but a limited AR display (it has a diagonal field of view of 52 degrees). This means that a user may be able to view the real-world elements without looking through the AR display. 
This work uses the HoloLens' 2 Optical See Through (OST) AR, as professional first responders need to not have their vision limited in any way and they require a compact device that minimizes interference with Personal Protective Equipment. Additionally, the resolution of many pass-through AR devices makes it difficult to see small details, making  contemporary pass-through AR devices not suitable. This may change in future. 
For development we used the Mixed Reality Toolkit (MRTK)~\cite{MRTK}, with the Unity Game Development Engine, which provides access to the spatial awareness mesh.
The HoloLens 2 spatial awareness mesh is continuously updated using 3D scans of the surrounding environment. It is essentially a simple 3D mesh provided without any texture coordinates. This means that if a standard texture is to be used, it is necessary to map from the 3D position to two-dimensional texture coordinates onto which the texture can be drawn. The mesh is typically drawn by applying a Unity material to the mesh to make it visible. 
The spatial awareness mesh must be built by the HoloLens 2 using its depth camera and it has limited extents. With our approach, being able to see the radiation requires that there is a spatial awareness mesh rendered near the source. 
In all practical training scenarios,  a radiation source will be on a surface, with walls and other structures nearby, that can be identified by the spatial awareness mesh. The only time there may not be a spatial awareness mesh around a source is  when the radiation source is  far from the HMD.  We address this by drawing a plane that is bill-boarded towards the user (i.e., always perpendicular to the camera), and texturing it with the radiation texture (but leaving it transparent otherwise), if the source is above a threshold distance. 
The plane fades as the trainee approaches, and the spatial awareness mesh around the source is filled in. This technique has been implemented but is not part of our visualization evaluation, as the room the experiment took place in can be fully described by the HoloLens 2 spatial awareness mesh. 
\subsection{Shader Overview}
Modern computer graphics approaches allow for custom programs, known as shaders, to be written and executed on a device's GPU. 
While there are some standard shaders supplied by the MRTK, the technical requirements for visualizing radiation levels in the mesh resulted in custom shaders being developed (more details are available in the supplemental materials). These shaders had to deliver the following functionality:
\begin{itemize}
	\item A point in the mesh must be colored based on the intensity of radiation at that position.
	\item Multiple sources can contribute to the radiation at any given point.
	\item To allow the users to see both the physical world and the virtual radiation, it must be possible to limit colors to a specific stencil shape leaving gaps in the coloring using a stencil.
	\item To help the user determine the best path to follow it must be possible to orient stencils in a specific direction.
\end{itemize}
\subsection{Implementation for Source Accumulation}

\begin{figure}[t]
     \centering
     \begin{subfigure}[b]{0.48\textwidth}
         \centering
         \includegraphics[trim={0 0 0 1cm},clip,width=0.9\textwidth]{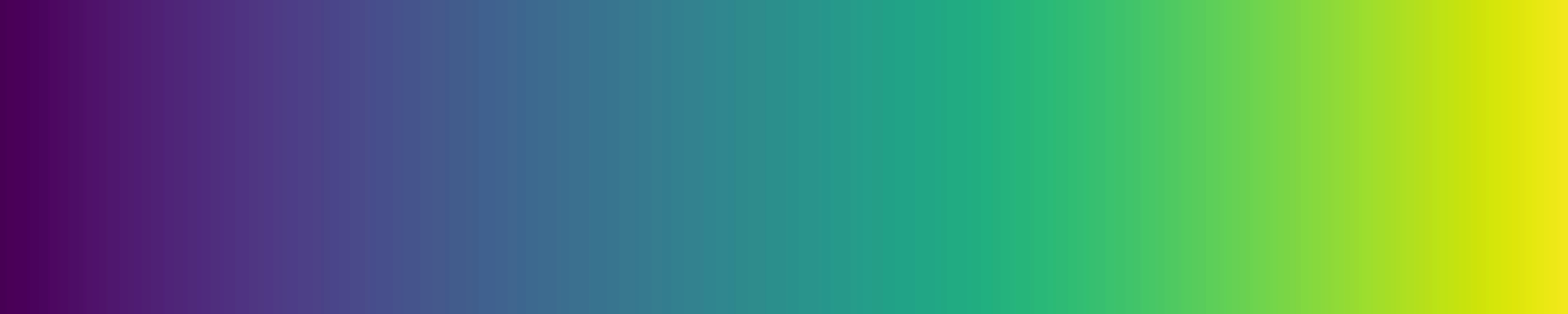}
         \label{fig:viridis}
     \end{subfigure}
     \par\medskip
     \begin{subfigure}[b]{0.48\textwidth}
         \centering
         \includegraphics[trim={0 0 0 1cm},clip,width=0.9\textwidth]{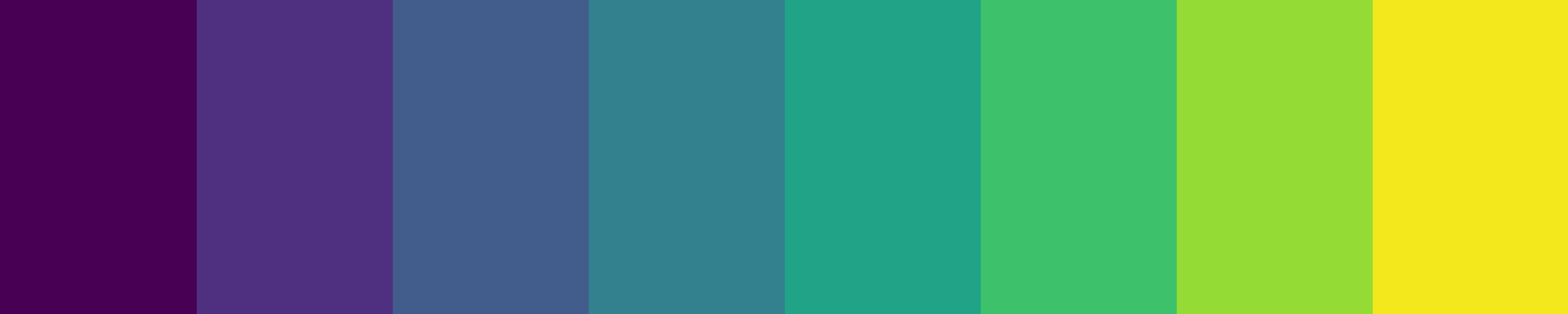}
         \label{fig:viridis8}
     \end{subfigure}    
        \caption{The color palettes sampled by the fragment shaders, the continuous on the top and the 8-color banded variant on the bottom. The yellow end of the scale indicates the highest radiation levels.}
        \Description{The full Viridis color palette  and a Viridis color palette  sampled into 8 bands of color}
        \label{fig:palettes}
\end{figure}

To calculate the intensity at a given point in the mesh, the shader needs to consider the position of a source as input as well as the mapping of the domain of the radiation intensity, defined by a lower and upper bound, to the output color range. The min and max value of the intensity domain are defined as shader inputs.  The output color range is defined by a texture (essentially a color bitmap), which contains the color scale. The input texture used for this work was generated for the Viridis scale. The only time another texture is used is when the color scale is to be banded, see Figure \ref{fig:palettes}.
The fragment position in world coordinates is interpolated from the world space position of the vertex.
The contribution from each radiation source at the fragment is calculated, based on the source positions which are defined as inputs. The intensity is calculated using exponential drop off (dose rate divided by the distance squared). The mapping between the final intensity and the color is done based on the maximum and minimum range inputs.
This fundamental approach for coloring a pixel based on radiation intensity does not require any texture coordinates (that is a mapping to coordinates of an input texture), as the only factor affecting the pixel is its distance from a source.
\subsection{Using a Stencil}
To avoid drawing every pixel with radiation intensity, we used a stencil-based approach. The radiation intensity will only be rendered for a pixel if it falls within a shape specified by the input stencil texture.
\begin{figure*}[t]
     \centering

     \begin{subfigure}[b]{0.2\textwidth}
         \centering
         \includegraphics[width=0.9\textwidth]{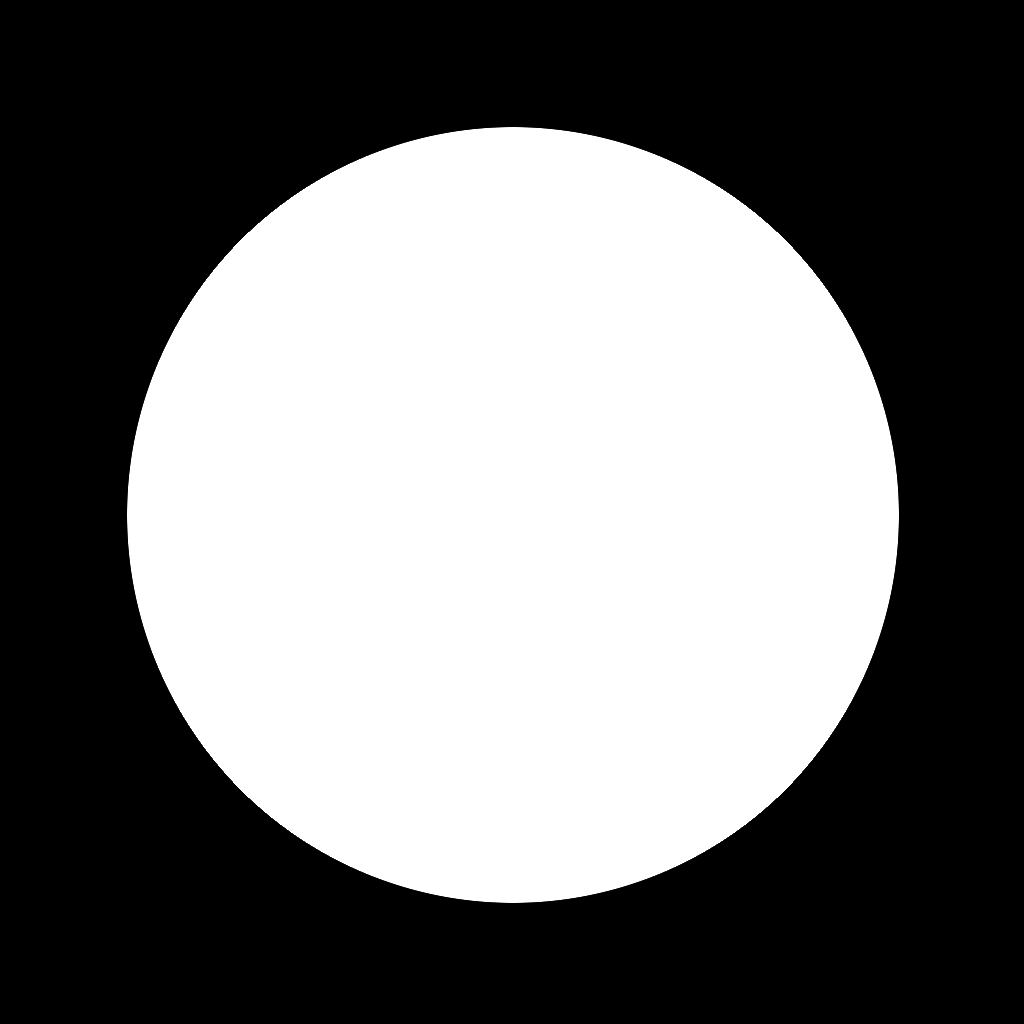}
         \label{fig:circle}
     \end{subfigure}
			\hfill
     \begin{subfigure}[b]{0.24\textwidth}
         \centering
         \includegraphics[width=0.9\textwidth]{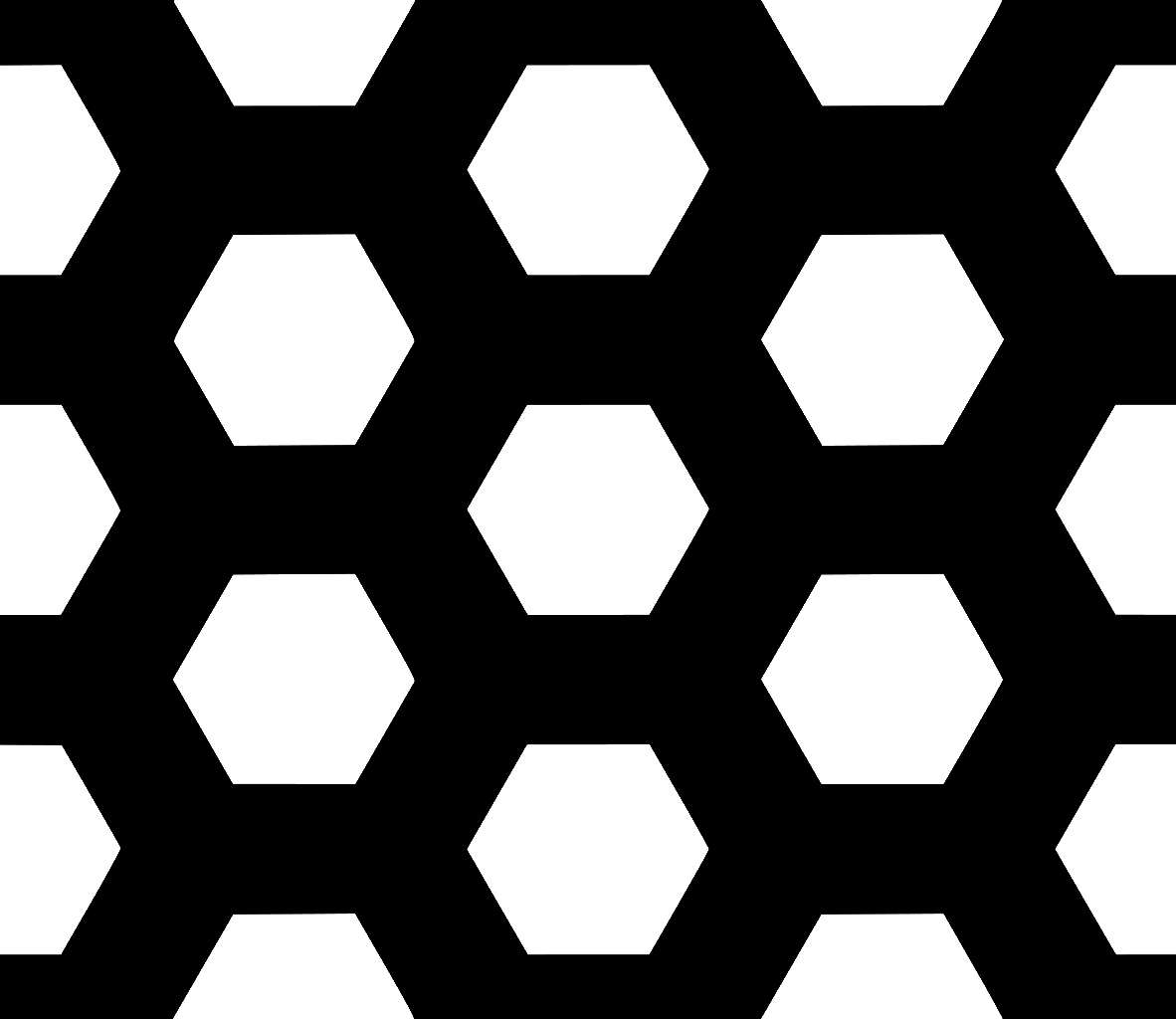}
         \label{fig:hex}
     \end{subfigure}    
		\hfill
		\begin{subfigure}[b]{0.2\textwidth}
         \centering
         \includegraphics[width=0.9\textwidth]{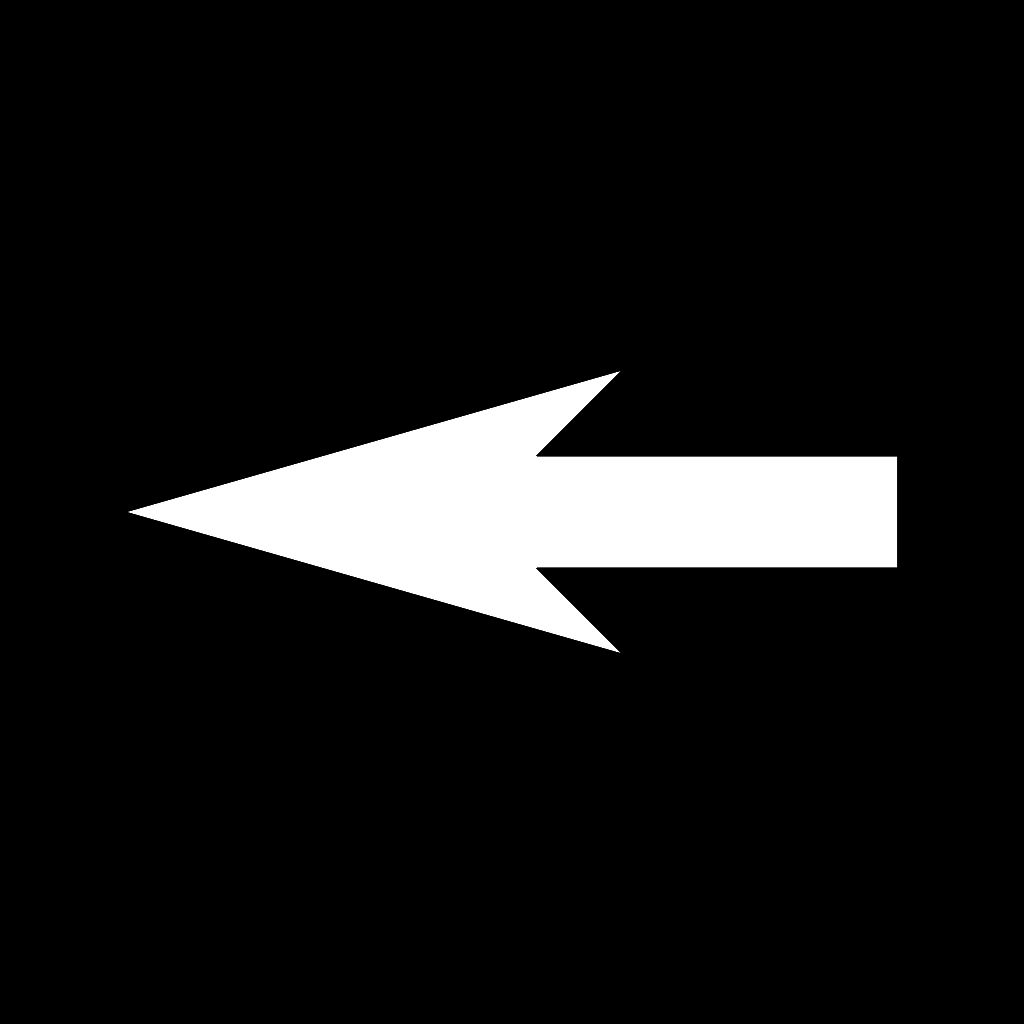}
         \label{fig:arrow}
     \end{subfigure}    
        \caption{The actual textures used for rendering each stencil, but transparency has been colored black. The patterns were tiled differently so the hexagons appeared visually similar in scale to the circles when rendered.}
        \Description{A white circle with black background, white hexagons with a black background, A white arrow on a black background pointing to the left}
        \label{fig:stencils}
\end{figure*}

We began with the simplest texture, a circle, however one potential issue with such a pattern is that the space between the circle texture is not consistent. 
Therefore, for a more regular spacing of the stencil we chose a hexagon, as the distance between adjacent hexagons is more consistent and hexagons vary little under rotation (although they are not fully rotationally invariant like circles).
One key issue to be faced was how to apply the stencil shape texture to the spatial awareness mesh. The spatial awareness mesh is consistently updated by the HoloLens hardware to reflect the world. It is dynamic messy data, the does not come with a neat set of texture coordinates that are easily looked up.
Therefore, we adopted the approach of tri-planar texture mapping.
Tri-planar texture mapping is a technique by which textures are rendered aligned with 3 planes (usually aligned with the coordinate system for practicality), and then the final pixel is colored based on a blend of the resulting textures.
To only show the radiation intensity within a specific stencil shape, we used the stencil buffer of the HoloLens graphics capabilities. The stencil buffer allows for fragments to only be drawn based on those specified in an initial pass, all fragments not drawn in this initial pass are not contributed to in subsequent rendering passes. This approach involves a rendering pass to the stencil buffer where the spatial awareness mesh is drawn using a stencil texture. Fragments for the final rendering pass are only considered where the stencil texture was drawn in the initial pass.
\subsection{Orientation}
The arrow stencil needs to be oriented in the correct direction, considering all 3 of the tri-planar textures.  To achieve this effect a rotation must be applied to each, rotating them to point away from the average source position, weighted by source intensity.
\subsection{Resulting Visualizations}
Different techniques were being evaluated, however there were some common characteristics. We have chosen the Viridis color scale \cite{Garnier2021} to be the basis of all stimuli.
This color scale has been recently evaluated as effective  \cite{Liu2018}  and is considered more resistant to red-green color blindness.
The change of color is exponential (as radiation values drop exponentially with distance). 
Other palates can easily be applied to our approach by changing the input texture. 
The focus of our work is how the visualization is rendered, not which color scale is chosen. Readers who wish to further explore the impact of colormap are referred to \cite{Reda2021} and \cite{Liu2018}.
The final list of visualizations can be seen in Table \ref{tab:vis}, and examples of the visualizations in Figure \ref{fig:visualizations}.
For details of the implementation of the visualizations see the supplemental materials.

\begin{table*}
	\caption{The visualization approaches.}
        \Description{A description of each of the visualization approaches applied.}
	\label{tab:vis}
		\begin{tabular}{c p{3cm} p{4cm} p{4.5cm}}
		\toprule
		Name & Description & Motivation & Experiment Hypothesis \\ 
		\midrule
		Continuous & This uses the full Viridis color scale & Radiation visualization using the spatial awareness mesh. & No hypothesis as this was the base condition for comparison. \\ 
		Banded & The color scale is divided into 8 color bands.  & Allowing experiment participants to clearly distinguish the changes in gradients. & H1: The banded visualization will outperform the continuous opaque visualization. \\
		Transparent & Full-scale color where the opacity of the color is set to 33\%. & To determine if transparency can effectively allow a user to perceive the real world along with the visualization. & H2: the transparent visualization will outperform the continuous opaque visualization. \\ 
		Circle & This uses the full color scale but with a circular stencil.  &  Use of a stencil to allow the user to see through the visualization. & H3: The stencil  visualization will outperform the continuous opaque visualization. \\
		Hex & Full color scale but with a hexagonal stencil. & Use of a stencil where the gaps  between the stencil items are consistently spaced. & H4: the hexagonal visualization will outperform the circular visualization. \\
		Arrow & Full colors scale with an arrow stencil that is oriented pointing away from sources. & Providing direct directional information will make the best path more clear.  & H5: the oriented arrow visualizations will outperform the continuous opaque visualization. \newline H6: the oriented arrow visualizations will outperform the other stencils. \\ 
		\bottomrule
\end{tabular} 

\end{table*}

\begin{figure*}[t]
     \centering

     \begin{subfigure}[b]{0.3\textwidth}
         \centering
         \includegraphics[width=\textwidth]{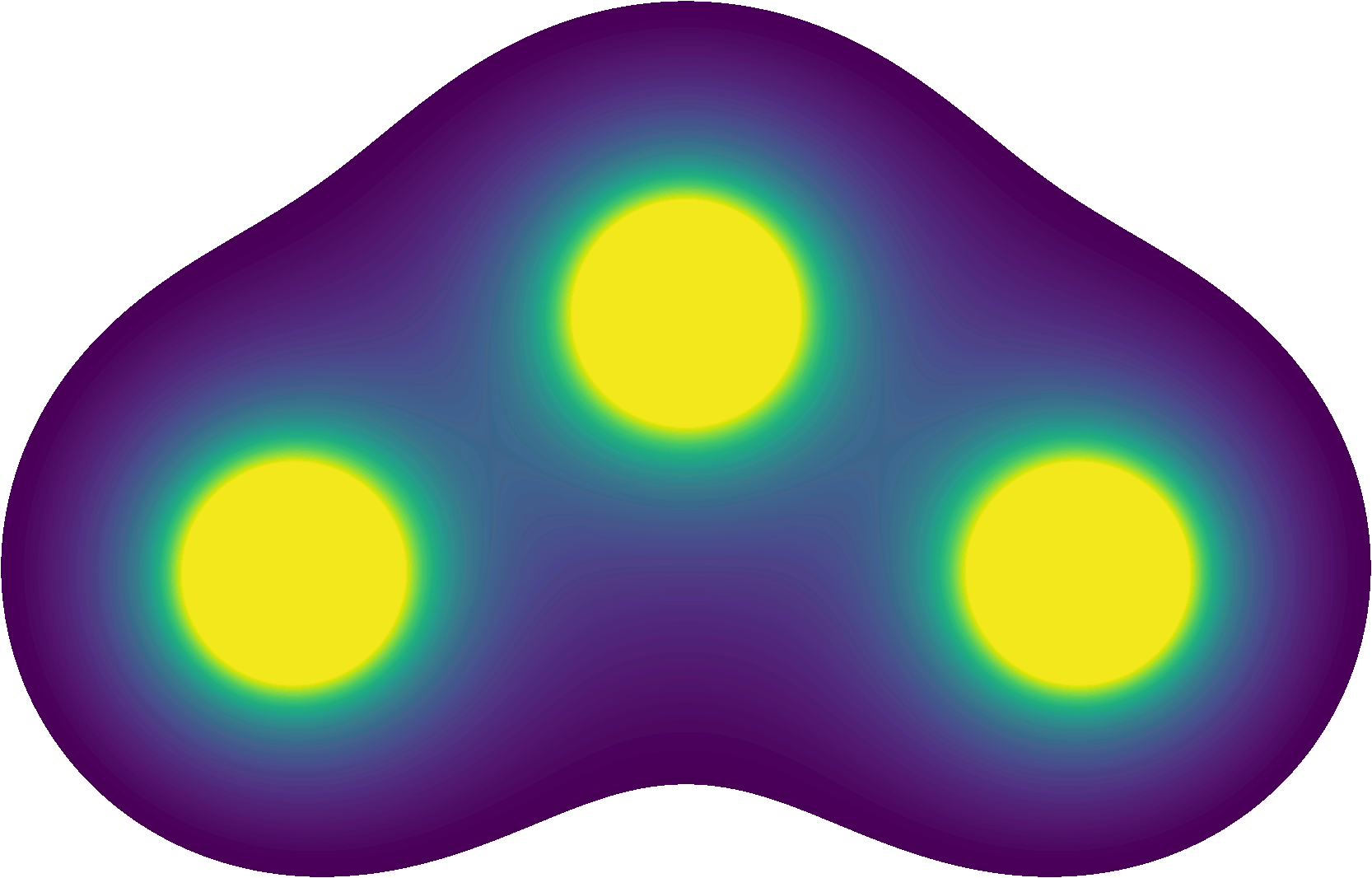}
				\Description{3 Radiation sources visualized using a continuous Viridis scale on a flat plane}
         \label{fig:cont}
     \end{subfigure}
			\hfill
     \begin{subfigure}[b]{0.3\textwidth}
         \centering
         \includegraphics[width=\textwidth]{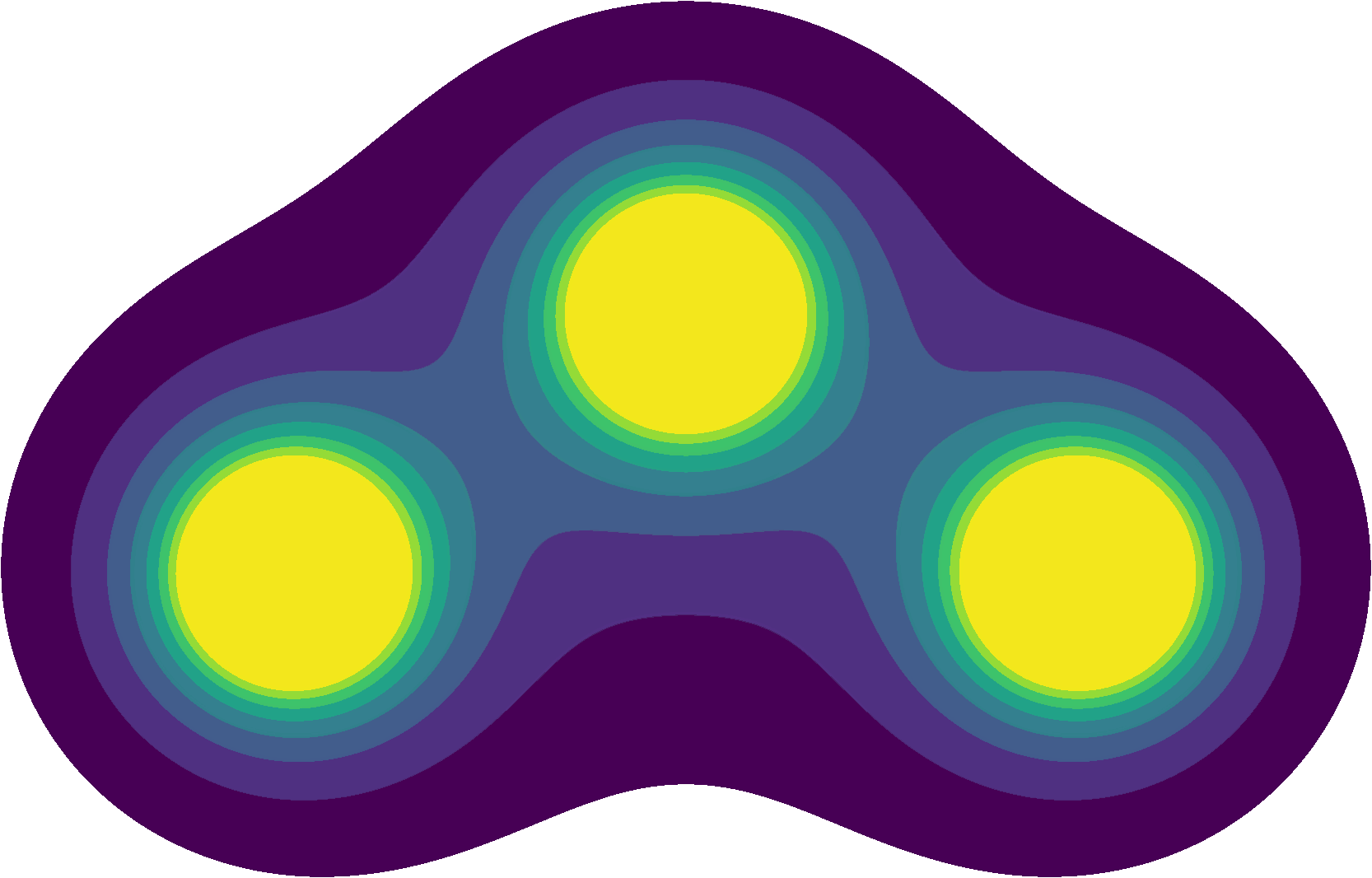}
				\Description{3 Radiation sources visualized using a banded Viridis scale on a flat plane}
         \label{fig:banded}
     \end{subfigure}    
		\hfill
		\begin{subfigure}[b]{0.3\textwidth}
         \centering
         \includegraphics[width=\textwidth]{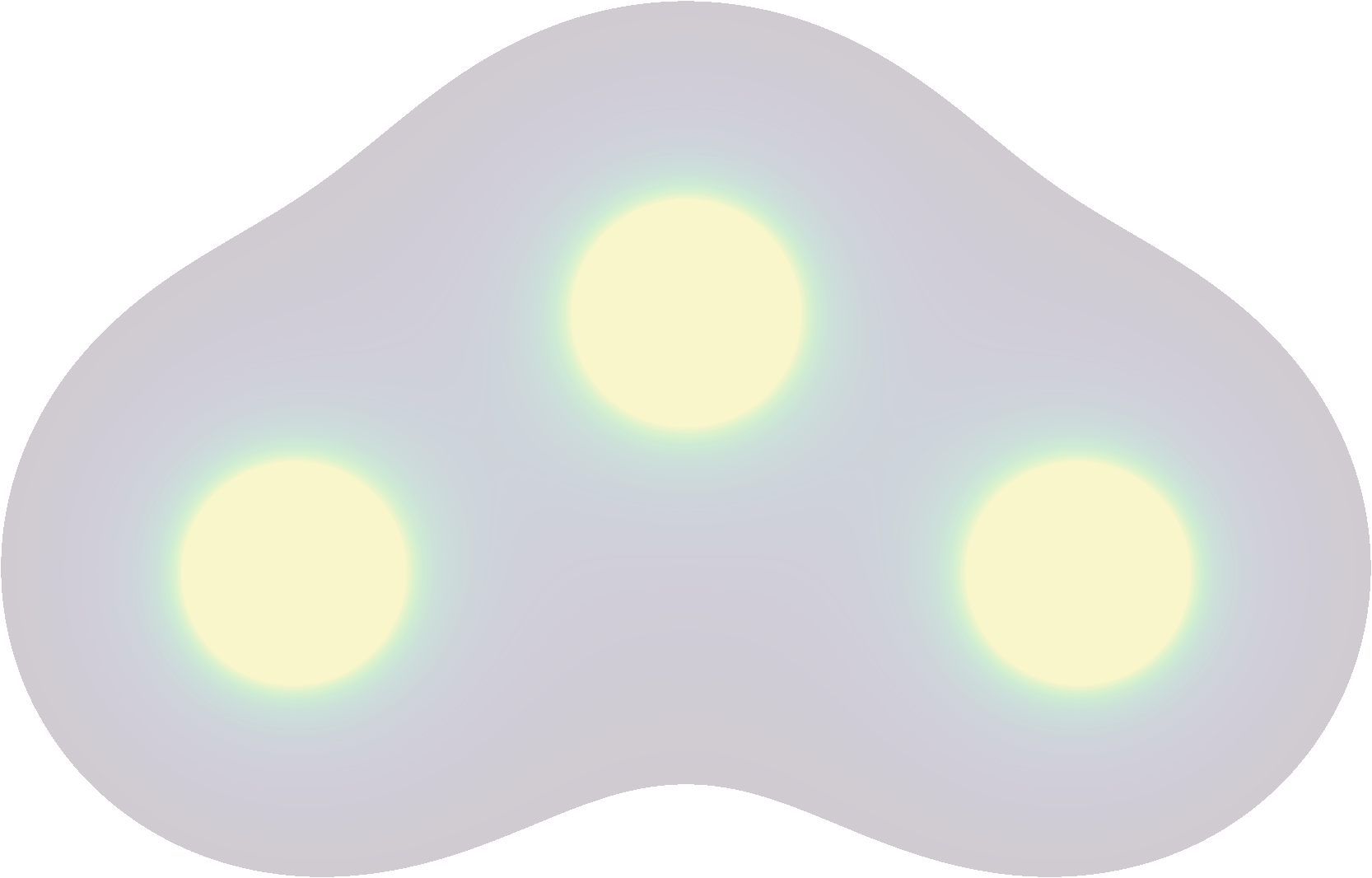}
				 \Description{3 Radiation sources visualized using a continuous semi-transparent Viridis scale on a flat plane}         
         \label{fig:trans}
     \end{subfigure}

		 \begin{subfigure}[b]{0.3\textwidth}
         \centering
         \includegraphics[width=\textwidth]{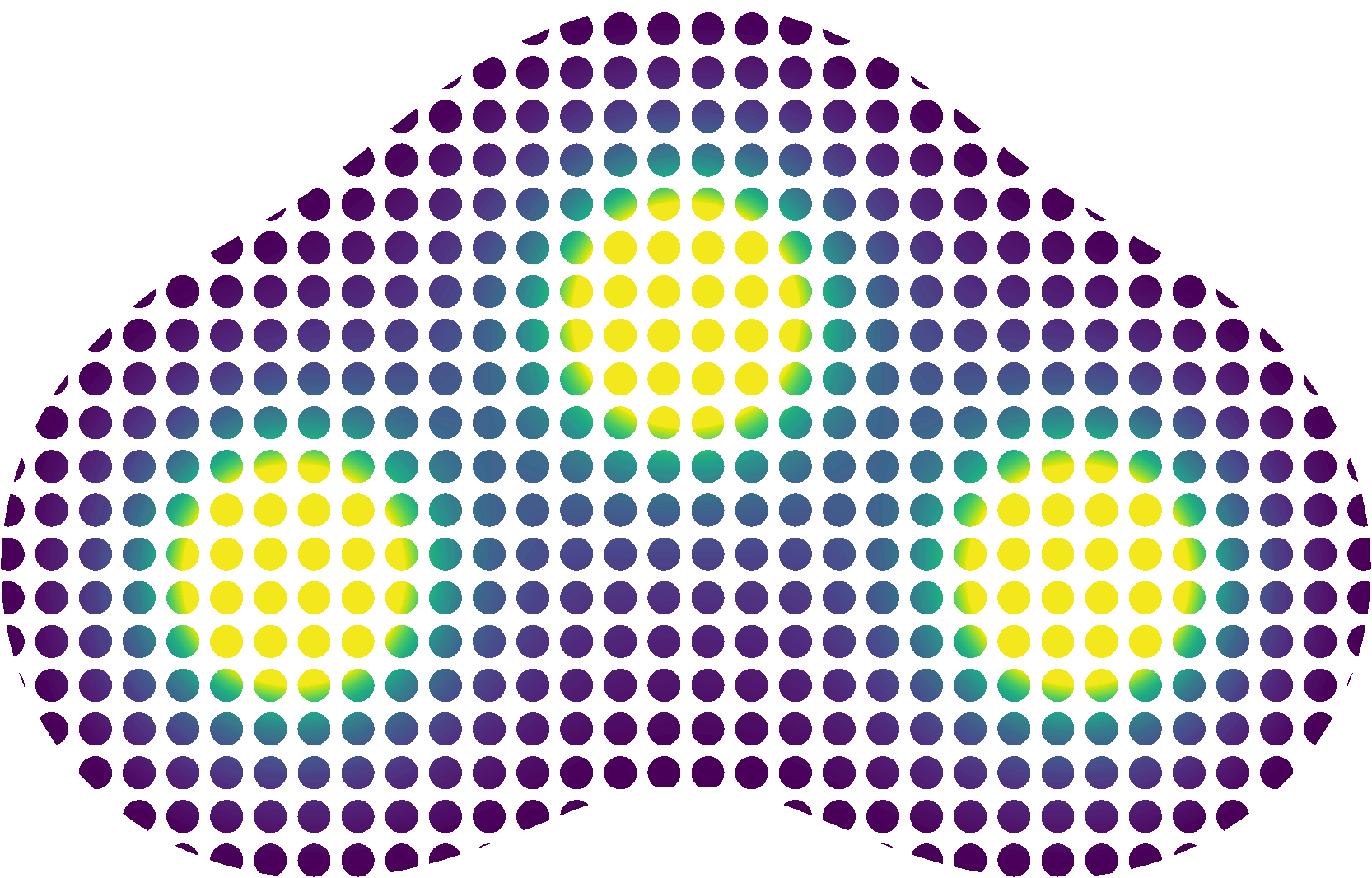}
				 \Description{3 Radiation sources visualized using a circle stencil and a continuous Viridis scale on a flat plane}   
         \label{fig:circle}		
     \end{subfigure}
			\hfill
     \begin{subfigure}[b]{0.3\textwidth}
         \centering
         \includegraphics[width=\textwidth]{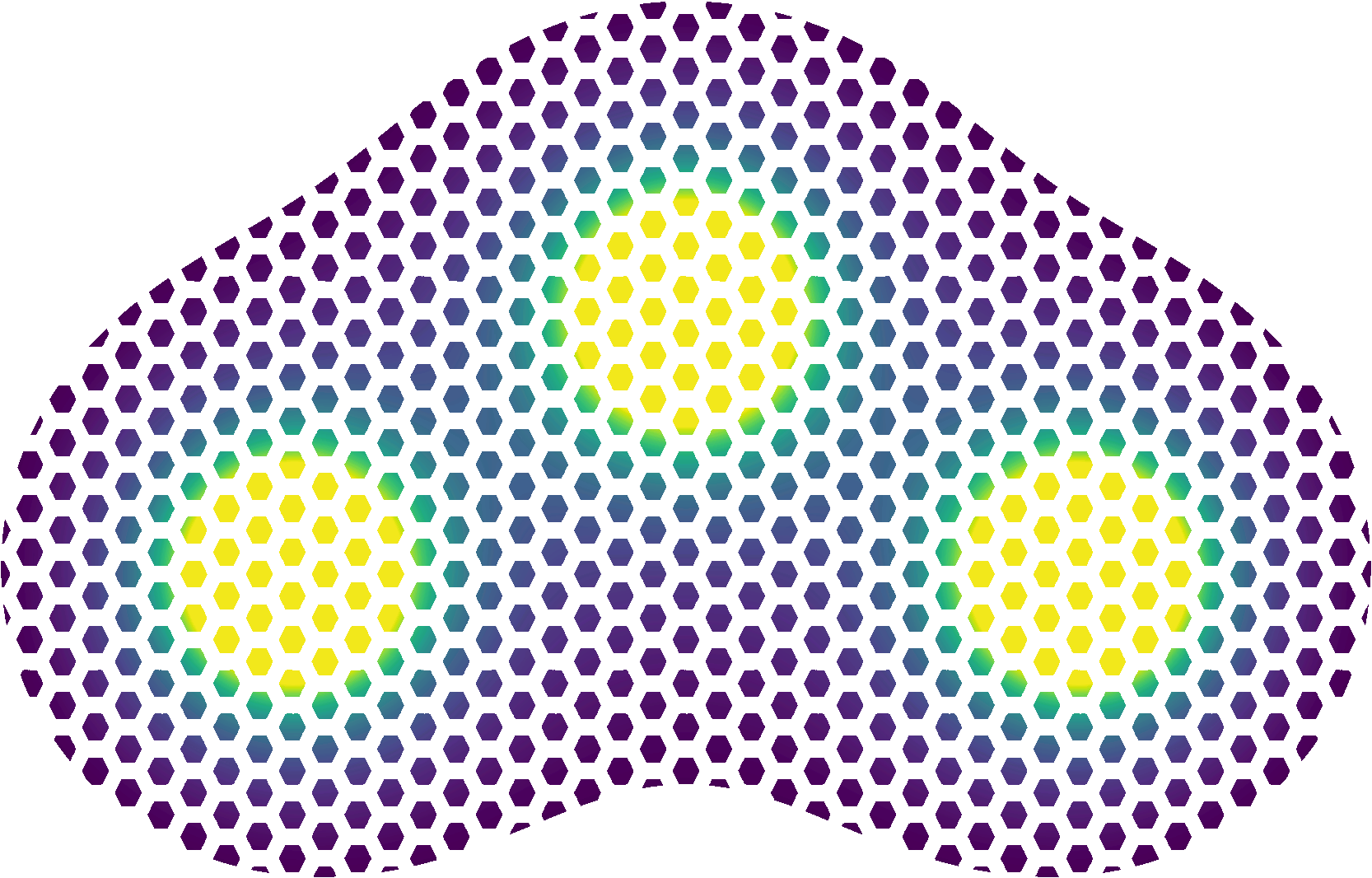}
				 \Description{3 Radiation sources visualized using a hexagon stencil and a continuous Viridis scale on a flat plane}  
         \label{fig:hex}
     \end{subfigure}    
		\hfill
		\begin{subfigure}[b]{0.3\textwidth}
         \centering
         \includegraphics[width=\textwidth]{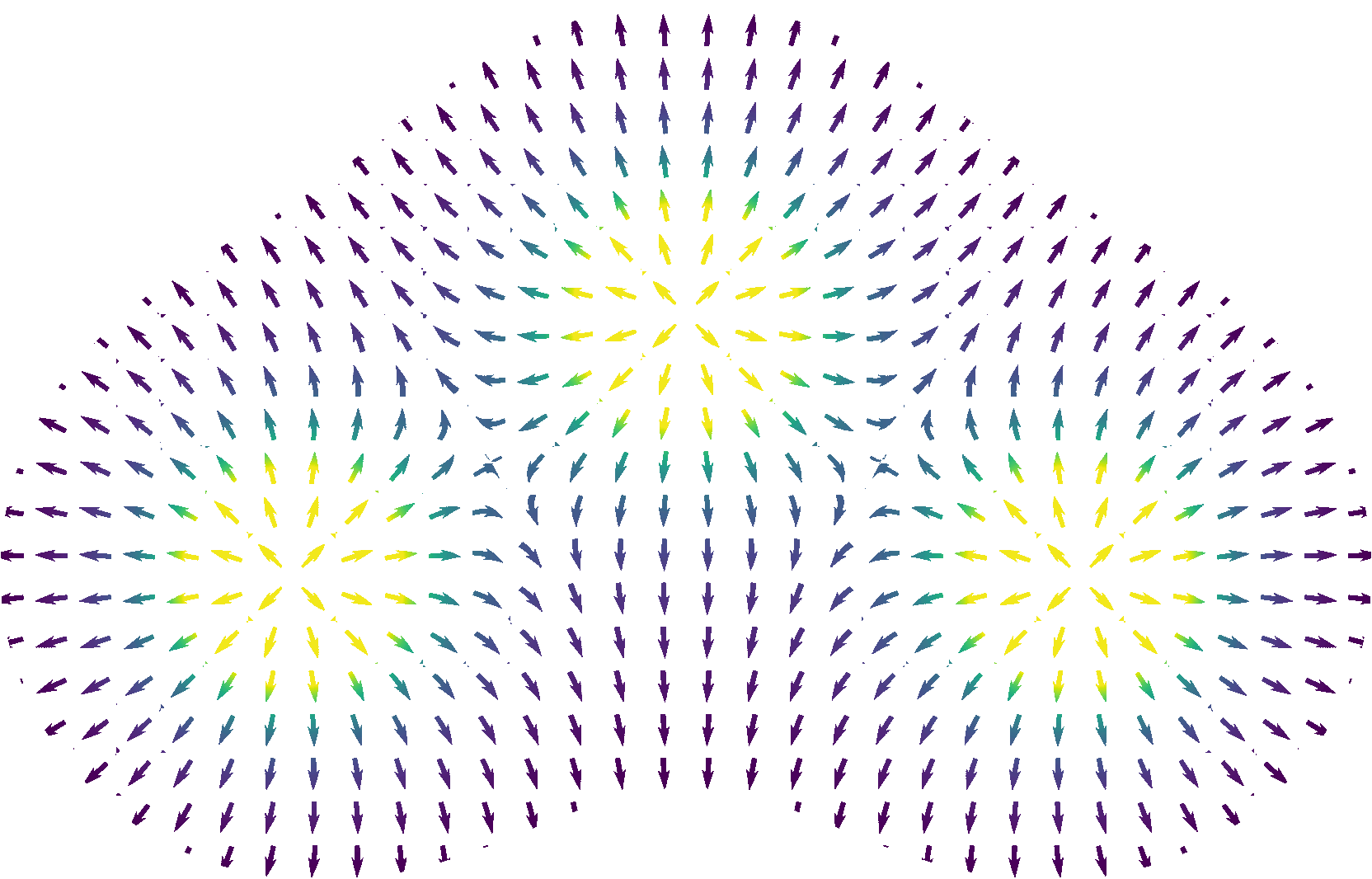}
				 \Description{3 Radiation sources visualized using an  arrow stencil and a continuous Viridis scale on a flat plane}   
         \label{fig:arrow}
     \end{subfigure}    		
        \caption{Illustrative examples of the visualizations on a flat plane, showing radiation from 3 sources.}
        \label{fig:visualizations}
\end{figure*}

\section{Visualization Evaluation}
We performed user experiments to determine the impact of each approach. Participants performed a simple task for each trial with a specific visual encoding, and we measured their exposure during each trial and the time taken to complete the task. We have also used a questionnaire and free-form responses to gauge their subjective feelings towards each visual encoding.
\subsection{Experiment Hypotheses}
We consider the continuous opaque visualization as the base technique, and the others e.g., using transparency, banding, a stencil-based approach and encoding direction as enhancements. The hypotheses related to each visualization can be seen in Table 1.
We propose an additional hypothesis that individual personal differences between participants, as measured by Visual Spatial Image ability (VSI) may impact on the results. VSI is essentially the ability of a person to imagine a 3D scene based on 2D, auditory or textual descriptions. We would expect people with higher VSI to more accurately build a mental model of the risks in the scene, therefore leading them to avoid risks more thus reducing their overall radiation dose absorption. VSI can be evaluated using a pre-experiment questionnaire such as the Visual Spatial Imagery set of Vorderer \emph{et al.} \cite{Vorderer2004}. Our additional hypothesis is as follows:
\begin{itemize}
	\item H7: Those who exhibit VSI ability will have a lower cumulative simulated radiation dose exposure.
\end{itemize}

\subsection{Experiment Setup}
\label{sec:setup}
\begin{figure}[h]
  \centering
  \includegraphics[width=0.8\linewidth]{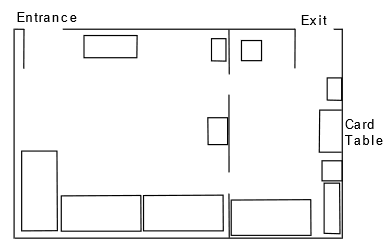}
  \caption{The layout of the experiment room.}
  \Description{A top down schematic  of the experiment room layout, including labels for the entrance, exit and card table location.}
  \label{fig:room}
\end{figure}

The Experiment took place in a room of 4.5M * 8.5M, with a partition in the middle, 4.5M from one end of the room. There was space at either side of the partition for the user to walk around it. At opposite ends of the room along one side wall are doors by which the room was entered and exited for the experiment (see Figure \ref{fig:room}).
 For each trial there were 3 virtual radiation sources in the room. For a specific trial, the user had to enter the room from one door wearing the HMD and perform the trial task in a brisk manner (not running), while keeping their exposure to the virtual radiation to a minimum. The experiment evaluated each of the visualization approaches described in section 3  across multiple trials. Each visualization type was used as a stimulus for a block of trials.
In addition to the 6 radiation visualizations, there were 5 different configurations (position-wise) of the 3 radiation sources in the room, referred to as scenes.  The aim was that the visualization would influence the user to determine which path to take to the table through the central partition. A specific trial consisted of a 1 of the 6 visualizations in 1 of the 5 scenes.  A sixth scene was used only for user training. The scene layouts are illustrated in the supplemental materials.
All radiation sources were of equivalent strength and the visualization was configured to have a radius of 2 meters for a single source. Radiation sources have dose rate measured in Sieverts (Sv) per hour at a 1 meter distance. This depends on the source substance and the amount of the material.  Each virtual source used for experiment had a dose rate of 1 millisievert per hour (mS/H) at a distance of 1 meter. However, the actual dose received was not the focus of the experiments or analysis, just whether the visualization type impacted performance. The range of the visualization can be configured for different values of inputs.

\subsection{Trial Task}
The trial task was designed to require that the user was able to understand both the radiation and the environment around them.
The task consists of the participant entering the experiment room, searching for an item and exiting the room during which time they must try to minimize their total radiation absorption.
In real-world scenarios, first responders will often search a scene to find evidence, such as a USB key or mobile phone.
We required a recognizable item of reasonably comparable size, so we chose a playing card (from a standard 52 card deck).
Having to select from a set of similar items means the participant needs to have some understanding of the physical world as well as the virtual.
The participants had to select the correct card from  amongst 26 other cards (a red suit and a black suit were used, to make them easier to distinguish) which are lying face up and presented in random order (See Figure \ref{fig:room} for a photo).
For each trial, the user is given a new card and asked to retrieve the matching one from the table. The purpose of retrieving the card is to demonstrate to the experimenter that the correct item was found. After each trial, the card was returned to the table in a different position from where it was retrieved. During a trial, the experimenter would remain outside of the room to avoid distracting the participant and to avoid being registered as part of the spatial awareness mesh. The cards on the table were fully re-randomized between trial blocks.

\begin{figure}[h]
  \centering
  \includegraphics[width=0.8\linewidth]{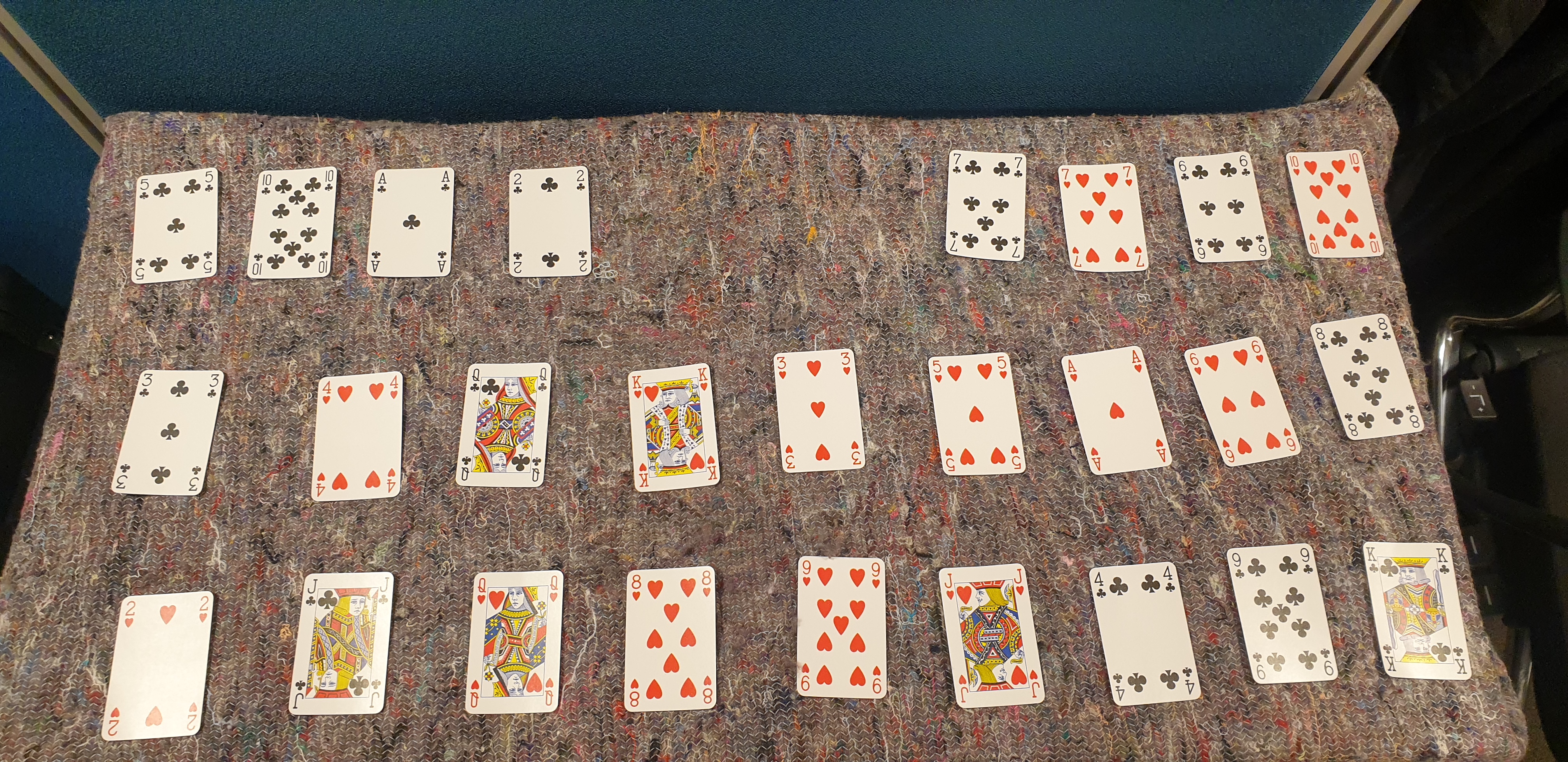}
  \caption{The table containing cards as used in the experiment.}
  \Description{A photograph of playing cards laid out on a table.}
\end{figure}

\subsection{Experiment Structure}
Participants were presented with consent forms and an experiment description sheet as well as an information sheet describing each of the visualizations (see supplemental materials). They were also given a verbal description of the experiment procedure. When ready, the participant began with a training block of trials to ensure they knew what to expect from the visualizations, ensure that the experiment task was clear and reduce any potential learning effect.
The training block consisted of 6 trials, one for each visualization, and each using a special training scene. The participants could explore the experiment space, and the experimenter was present with them in the room for these trials to answer any questions. The order of visualizations was randomized for the training block.
The experiment was divided into 6 blocks one for each visualization.  At the end of a block, the participant answered a questionnaire concerning that specific visualization (described in more detail in Section 5).  The order of blocks was randomized for each participant. Each block consisted of 5 trials, one for each of the five configurations of sources (referred to as scenes). The order of the scenes in each block were also randomized. This meant the that in addition to the training trials each participant performed 30 trials.

\section{Evaluation Approach}
In this section we describe the materials and metrics used to evaluate the visualizations.
\subsection{Pre-Experiment Questionnaire}

\begin{table*}
	\caption{The pre-experiment questionnaire.}
        \Description{The pre-experiment questionnaire with questions from the MEC-SPQ Visual Spatial Imagery (VSI) 4 item scale [Vorderer et al. 2004]}
	\label{tab:preExp}
		\begin{tabular}{c c p{6cm} p{4cm}}
		\toprule
		\# & Source & Question Text & Answer Format \\ 
		\midrule
		  1 & MEC SPQ \cite{Vorderer2004} & When someone shows me a blueprint, I am able to imagine the space easily. & 7-point Likert Scale \\
		  2 &  & It's easy for me to negotiate a space in my mind without actually being there.  & \\ 
			3 &  & When I read a text, I can usually easily imagine the arrangement of the objects described. & \\ 
			4 &  & When someone describes a space to me, it’s usually very easy for me to imagine it clearly. & \\ 
		\bottomrule
		\end{tabular} 
\end{table*}

Prior to starting the experiment, we asked the user questions from the MEC-SPQ Visual Spatial Imagery (VSI) 4 item scale \cite{Vorderer2004}\, see Table 1.  
\subsection{Quantitative Metrics}
For each trial, the time elapsed (seconds) and user position were logged. The trial timer was controlled by the experimenter using a custom application on an Android device, clicking on a trial start button when the participant entered the room and stop when they exited the room having successfully completed the task. The data was further cleaned by removing data points from beyond the threshold of the entrance and exit doorways. The following error metrics were then calculated based on the time and position data.
\begin{enumerate}
	\item Cumulative radiation exposure during the trial (Sieverts)
	\item Mean dose rate received (Sieverts per Hour)
	\item Mean distance to the nearest source (Meters)
	\item Maximum dose rate received (Sieverts per hour)
	\item Time in proximity (1.5 meters) to the card table (Seconds)
\end{enumerate}

\subsection{Visualization Questionnaire (Subjective Data)}
The goal was to understand which techniques were more effective, in terms of maintaining awareness of both the real and virtual worlds. We also wanted to better understand the situational awareness aspect of each visualization approach. Therefore, we also included questions from existing spatial awareness evaluation surveys. We took questions from the MEC SPQ \cite{Vorderer2004} specifically the Spatial Situation Model (SSM) 4 item scale, and SART (the situational Awareness  Rating Technique) \cite{Taylor1990}, see Table 2. SART is mainly drawn from aeronautics and highly dynamic environments, as our scenes are static these aspects were removed. As such, we assess individual ratings rather than the overall score provided by SART.

\begin{table*}
	\caption{Questionnaire taken after each visualization block.}
        \Description{The questionnaire taken after each visualization block, featuring questions from the MEC Spatial Presence Questionnaire [Vorderer et al. 2004] and  SART (situation Awareness) [Taylor 1990] questionnaires, along with new questions}
	\label{tab:questionnaire}
		\begin{tabular}{c c p{9cm} p{2cm}}
		\toprule
		\# & Source & Question Text & Answer Format \\ 
		\midrule
		1   & This Paper & I could maintain suitable awareness of virtual radiation sources. & 7-point Likert Scale\\ 
		2  &  & I could maintain suitable awareness of my physical environment. &  \\ 
		3	 &  &  The balance of visibility between the physical and virtual was appropriate.& \\ 
		\midrule
		4	 & MEC SPQ \cite{Vorderer2004}  &I was able to imagine the arrangement of the spaces presented in the augmented reality experience very well.  & 7-point Likert Scale\\ 
		5	 &  & I had a precise idea of the spatial surroundings presented in the augmented reality experience. & \\
		6	 &  & I was able to make a good estimate of the size of the presented space. & \\
		7	 &  & Even now, I still have a concrete mental image of the spatial environment.  & \\
		\midrule
		8 & SART \cite{Taylor1990}  &Complexity of the situation: How complex is the situation, is it complex with many interrelated components (high) or is it simple and straightforward (low)?  & Number in range  1 to 7 \\
		9	&  & Arousal: How aroused are you in the situation, are you alert and ready for activity (High) or do you have a low degree of alertness?  & \\
		10	 & &Concentration of Attention: How much are concentrating on the situation? Are you concentrating on many aspects of the situation (High) or are you focused on only one (Low)?& \\
		11	 &  &Division of Attention: How much is your attention divided in the situation? Are you concentrating on many aspects of the situation (High) or focused on only one (Low)?& \\
		12	 &  &Spare mental Capacity: How much mental capacity do you have to spare in the situation? Do you have sufficient to attend to many variables (High) or nothing to spare at all (Low)?&   \\
		13	 &  &Information Quantity: How much Information have you gained about the situation? Have you received and understood a great deal of knowledge (High) or very little (Low)?&   \\
		\bottomrule
		\end{tabular} 
\end{table*}

\subsection{Visualization ranking}
Finally, the participants were asked to rank the visualizations sorting them from best to worst in their opinion, and to provide free text explaining their choice, and adding any further information they thought was relevant.
\section{Results}
Twenty-five adult participants took part, with  19 identifying as male and 6 as female, and  the majority had a computer science research background (see supplemental materials for further demographic information). One participant was removed from the quantitative analysis due to missing data for a small subset of trials, meaning they could not be included in the within-subjects analysis. However, the survey responses and ranking data were kept as they were complete with no errors. 

\subsection{Quantitative Evaluation}
\begin{figure*}[t]
     \centering
				\begin{subfigure}[b]{0.5\textwidth}
         \centering
         \includegraphics[width=\textwidth]{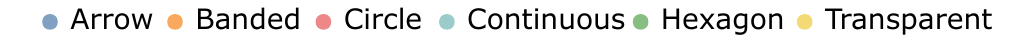}				 
				 \Description{The color legend for the results charts}
         \label{fig:result_time}
     \end{subfigure}
     \\
     \begin{subfigure}[b]{0.28\textwidth}
         \centering
         \includegraphics[width=0.9\textwidth]{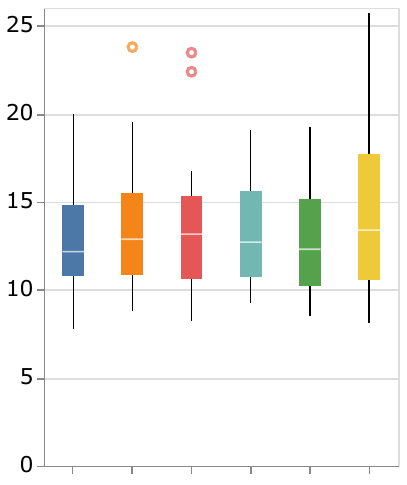}
				 \caption{Time(s)}
				 \Description{A Tukey boxplot showing the experiment results for time}
         \label{fig:result_time}
     \end{subfigure}
			\hfill
     \begin{subfigure}[b]{0.28\textwidth}
         \centering
         \includegraphics[width=0.9\textwidth]{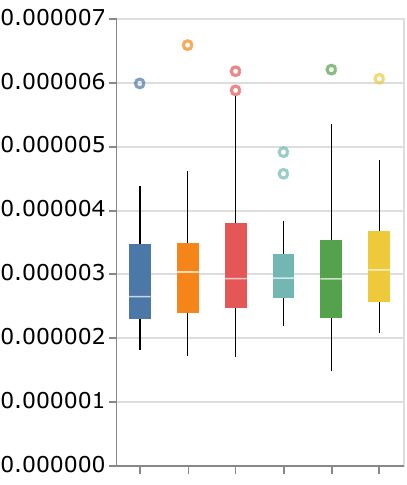}
				 \caption{Cumulative radiation (Sv)}
				 \Description{A Tukey boxplot showing the experiment results for cumulative radiation}
         \label{fig:result_cumul_rad}
     \end{subfigure}    
		\hfill
		\begin{subfigure}[b]{0.28\textwidth}
         \centering
         \includegraphics[width=0.9\textwidth]{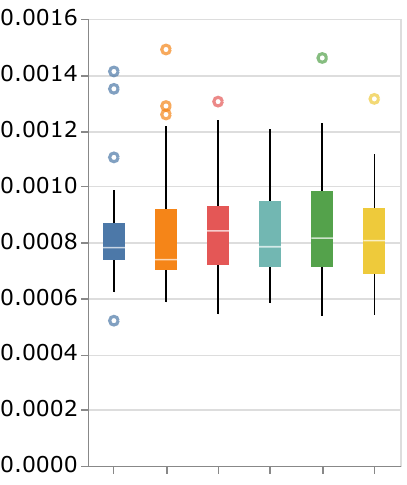}
				 \caption{Mean Dose Rate (Sv per hour)}
				 \Description{A Tukey boxplot showing the experiment results for mean dose rate}
         \label{fig:result_mean_dose}
     \end{subfigure}

		 \begin{subfigure}[b]{0.28\textwidth}
         \centering
         \includegraphics[width=0.9\textwidth]{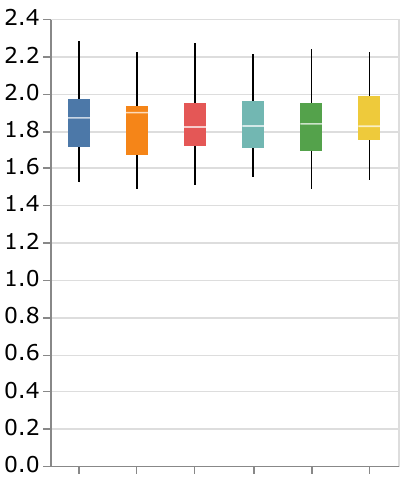}
				 \caption{Mean nearest source distance (M) }
				 \Description{A Tukey boxplot showing the experiment results for mean nearest source distance}
         \label{fig:result_mean_dist}
     \end{subfigure}  
			\hfill
     \begin{subfigure}[b]{0.28\textwidth}
         \centering
         \includegraphics[width=0.9\textwidth]{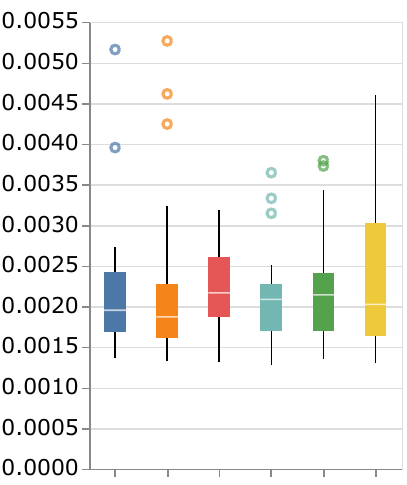}
				 \caption{Max dose rate (Sv per hour)}
				 \Description{A Tukey boxplot showing the experiment results for Max dose rate (Sv per hour)}
         \label{fig:result_max_dose}
     \end{subfigure}      
		\hfill
		\begin{subfigure}[b]{0.28\textwidth}
         \centering
         \includegraphics[width=0.9\textwidth]{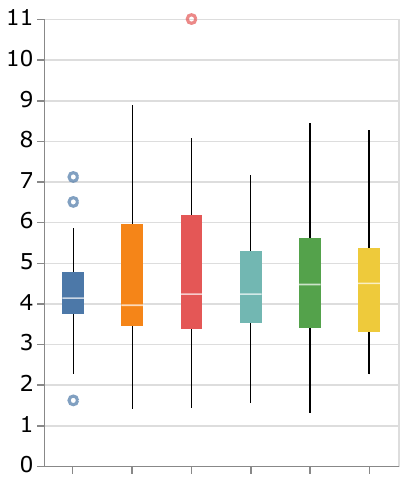}
				 \caption{Time (s) in proximity to table}
				 \Description{A Tukey boxplot showing the experiment results for Time in proximity to table}
         \label{fig:result_cumul_rad}
     \end{subfigure}      		
        \caption{Tukey boxplots for each of the metrics used for evaluation, with visualization approach being indicated by color.}
        \Description{Multiple Charts showing the results for each visualization}
        \label{fig:results}
\end{figure*}

The scores for each participant can be seen in the boxplots of Figure \ref{fig:results}. Each of the 6 metrics was tested for normalcy of distribution across all trials using a Shapiro-Wilks test. The only metric which was normal was mean nearest source distance. Therefore, all other metrics were tested for significance using a Friedman test, while mean nearest source distance was tested using ANOVA. The tests showed that there is no significant difference (p < 0.05) between the visualizations in terms of any metric. For all metrics tested using the non-parametric Friedman test, the effect size was determined to be small, using Kendall’s W. Therefore, none of our hypotheses can be accepted with respect to performance.

 \subsection{Visualization Questionnaire}
Due to the ordinal nature of the Likert scale data, we applied a Friedman test to the results.
The only question showing significant differences in answers was Q01, focusing on awareness for virtual radiation sources (Friedman test: p < 0.05). However, a Wilcox test showed that this potential significant difference was only between the circle and banded visualizations, and when a Bonferroni correction was applied the significance was maintained. However, this significant difference does not relate directly to any of our hypothesis, therefore none of our hypotheses can be accepted with respect to any of questions from the evaluation questionnaire.
\begin{figure*}[h]
  \centering
  \includegraphics[width=1.0\linewidth]{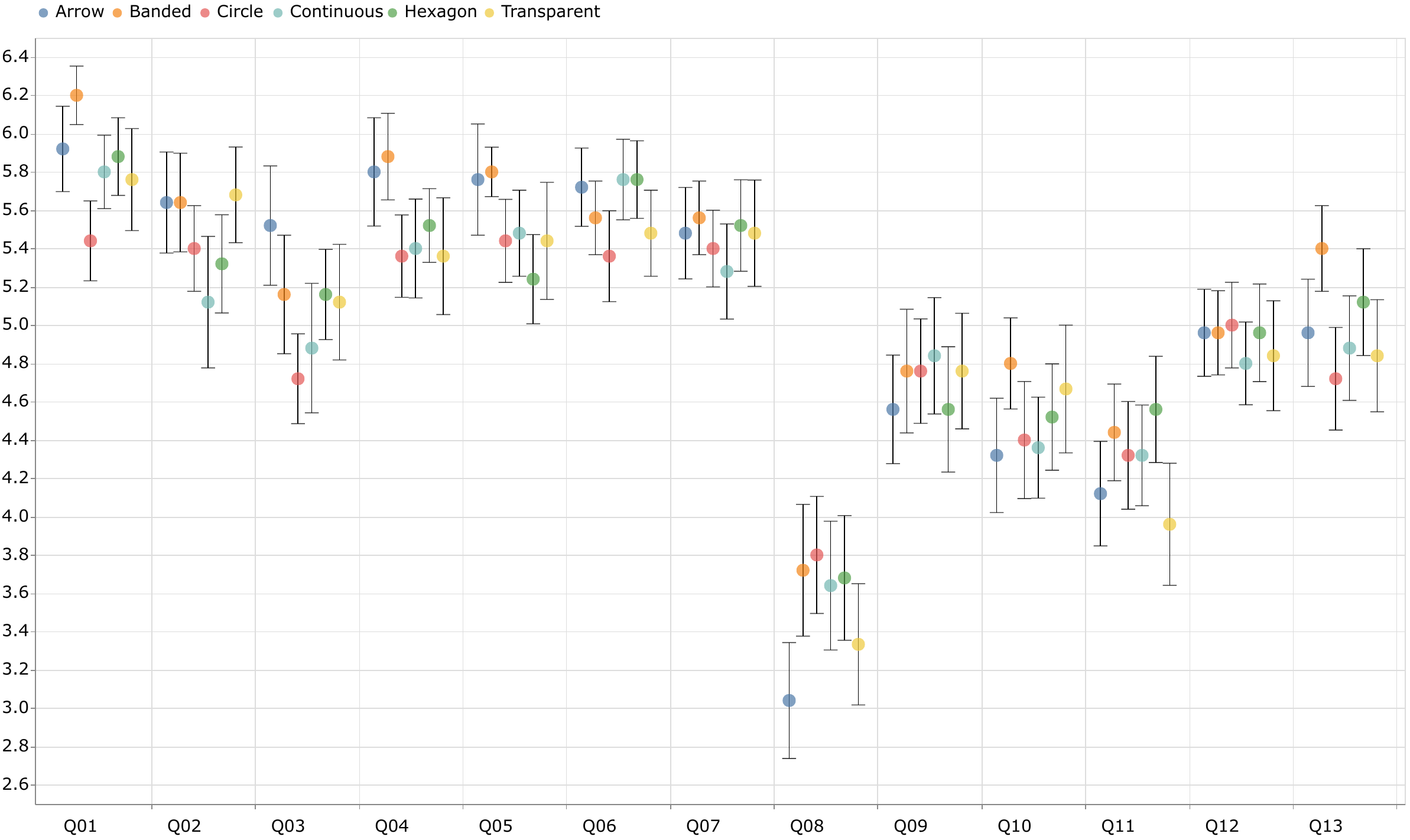}
  \caption{The results of the post trial block questionnaires (bars indicate standard error). The questions Q01 to Q07 were all asked on a 7-point Likert scale ranging from strongly disagree (a value of 1) to strongly agree (a value of 7). The remaining questions were answered on a numeric scale of 1 to 7.}
  \Description{A  bar chart show the rankings  of user preference with Arrow and Banded tied for highest  rank}
	\label{fig:questionnaire_results}
\end{figure*}

\subsection{Visualization Ranking}
\begin{figure}[h]
  \centering
  \includegraphics[width=0.9\linewidth]{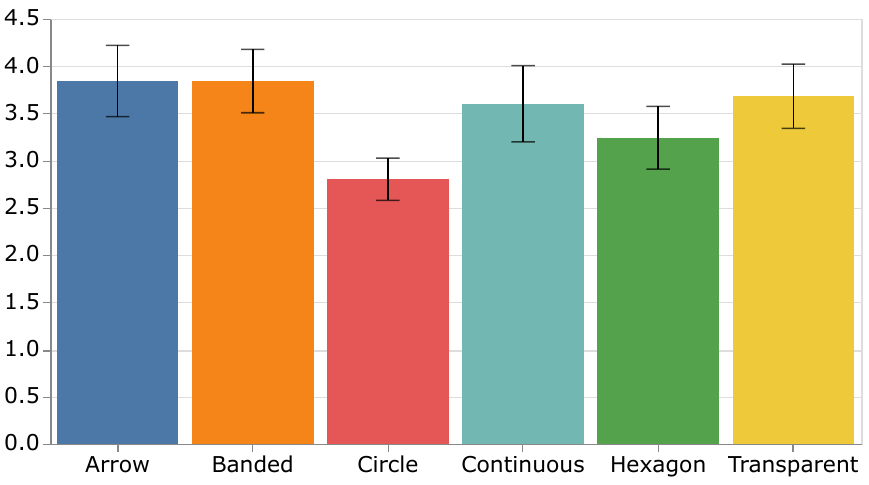}
  \caption{Average ranking for each visualization with standard error (1 = bottom ranked, 6 = top ranked)}
  \Description{A chart  showing the average rankings of the questionnaire for each visualization}
	\label{fig:ranking_results}
\end{figure}

\begin{figure*}[t]
     \centering
				
     \begin{subfigure}[b]{0.28\textwidth}
         \centering
         \includegraphics[width=\textwidth]{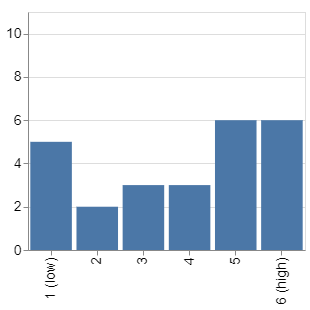}
				 \caption{Arrow}
				 \Description{A bar chart showing a binary distribution of rankings for the arrow visualization}
         \label{fig:arrow_ranking}
     \end{subfigure}
			\hfill
     \begin{subfigure}[b]{0.28\textwidth}
         \centering
         \includegraphics[width=\textwidth]{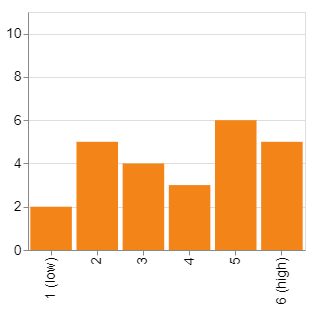}
				 \caption{Banded}
				 \Description{A bar chart showing a  distribution of rankings for the banded visualization}
         \label{fig:babded_ranking}
     \end{subfigure}    
		\hfill
		\begin{subfigure}[b]{0.28\textwidth}
         \centering
         \includegraphics[width=\textwidth]{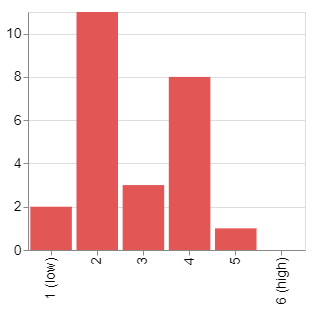}
				 \caption{Circle}
				 \Description{A bar chart showing a  distribution of rankings for the circle visualization.}
         \label{fig:circle_ranking}
     \end{subfigure}

		 \begin{subfigure}[b]{0.28\textwidth}
         \centering
         \includegraphics[width=\textwidth]{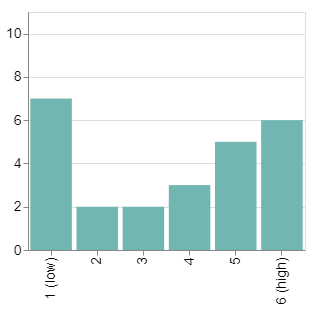}
				 \caption{Continuous}
				 \Description{A bar chart showing a  binary distribution of rankings for the continuous visualization.}
         \label{fig:result_mean_dist}
     \end{subfigure}  
			\hfill
     \begin{subfigure}[b]{0.28\textwidth}
         \centering
         \includegraphics[width=\textwidth]{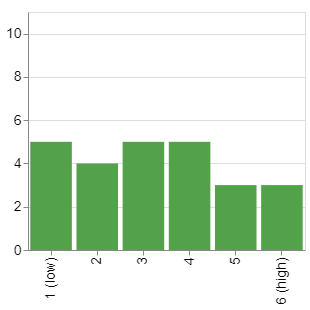}
				 \caption{Hex}
				 \Description{A bar chart showing a  distribution of rankings for the hex visualization.}
         \label{fig:result_max_dose}
     \end{subfigure}      
		\hfill
		\begin{subfigure}[b]{0.28\textwidth}
         \centering
         \includegraphics[width=\textwidth]{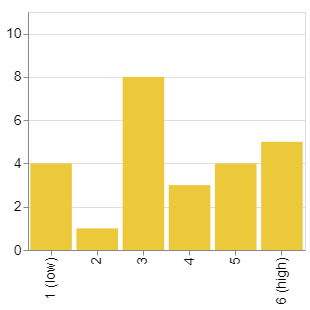}
				 \caption{Transparent}
				 \Description{A bar chart showing a  distribution of rankings for the circle visualization.}
         \label{fig:result_cumul_rad}
     \end{subfigure}      		
        \caption{The count of rankings for each visualization (1 = bottom ranked, 6 = top ranked).  }
         \Description{Multiple Charts showing the count of rankings for each visualization (1 = bottom ranked, 6 = top ranked).}
        \label{fig:ranking_distributions}
\end{figure*}

We analyzed the rankings, as can be seen in Figures \ref{fig:ranking_results} and \ref{fig:ranking_distributions}, and there were no significant differences across the users’ preferences, using a Friedman Test (p<0.05).

\subsection{Choice of Path}
As discussed in section \ref{sec:setup} and shown in the supplemental materials, for 4 of the scenes there was one side on the central room partition that exposed the user to more radiation. In order to consider if visualization had an impact on path choice, we counted the number of times the path of highest exposure was taken. Of the 576 trials logged for a scene with a clear path of higher exposure, a participant chose it 71 times. There were no significant differences between the visualizations.

\section{Discussion}
The lack of significant results in the user evaluation is somewhat surprising, however the results are interesting in and of themselves, given the differences shown in the boxplots of Figure \ref{fig:results}. Some insight can be gained by examining the distribution of rankings of the visualizations, as well as looking at the qualitative comments offered by the users.
As noted earlier, the role of visualization here was to influence human movement primarily through improving their situation awareness (perception, comprehension, planning),  while maintaining a suitable balance between the real and the virtual. The goal is for the visualizations to encourage the user to take the safest path.
In other words,  minimizing radiation exposure, which is a product of time and proximity to sources. We took a null hypothesis significance testing approach to analyzing our data. 
Such an approach is frequently used in visualization evaluation, in the context of determining  whether a visualization helps you answer a specifies question more correctly or more quickly. 
For this experiment we were trying to determine whether the visualization influenced the participants movement, while also not impeding them at the card finding task. It is possible that these two goals were somewhat confounding factors. While it is too early to provide a clear indication, it could be that further work is required to understand more clearly how human movement, not just understanding of the data, can be influenced by visualization techniques. This perhaps requires a greater analysis of the role of visualization on egocentric understanding e.g., the understanding of the distance between the observer and the stimulus.

\subsection{Performance Data}
In the Tukey box plots of Figure \ref{fig:results}, we can see that in terms of cumulative radiation the arrow visualization performs well relative to other approaches with a lower median value, a reasonable range of values, and only a single outlier. This is reflected in the lower median time and median nearest source distance. The arrows visualization approach resulted in the lowest average radiation dose (2.97 micro-Sieverts), and lowest average time (12.84 seconds).  The arrow visualization also resulted in a smaller distribution of values for time in proximity to the table. This is understandable, as the arrow stencil provides the least amount of spatial awareness coverage of all the visualizations. Unfortunately, the differences are not large enough for statistical significance. Looking at the mean nearest distance scores (Figure \ref{fig:result_mean_dist}), participants rarely spent time within 1.5 meters of the sources. As dose rate drops off exponentially with distance, this means that time was the largest contributing factor. All the visualizations, even the base continuous one, were effective at keeping the participants away from the sources. While there were differences in time, they were not enough for the cumulative radiation value to be affected to a significant degree.

\subsection{Questionnaire Results Discussion}
The results for the qualitative data were somewhat consistent with the results for the performance data. As can be seen in Figure \ref{fig:questionnaire_results}, with respect to the first 3 questions concerning the balance for the real and virtual, the responses of the participants do not diverge hugely from the performance analysis. On average the circle visualization is ranked lowest, with arrow and banded being ranked higher.  Q02, concerning maintaining suitable awareness of my physical environment, is the only instance of transparent visualization having the highest average score (5.68).
For MEC SPQ, the results generally follow this pattern, although it is notable that the hexagon based visualization comes near the top of the pack concerning estimating the size of the space and maintaining a concrete metal image of the space afterwards. This may indicate a different preference for a visualization approach depending on the type of the task, however the lack of a significant result means much further research and experimentation are needed to tease out the scale and nature of any such preference.

\subsection{Ranking Results}
 The results for quantitative data and ranking data were also somewhat aligned. The arrows visualization was joint highest rated in terms of preference along with the banded visualization (tied for first, with an average ranking score of 3.84, see Figure \ref{fig:ranking_results}).  
 Once again there were no significant differences, however we can see in Figure \ref{fig:ranking_distributions} that two of the more popular visualizations (Arrow and Continuous) have a binary distribution, many people giving it a top (or close) rank and a bottom rank. It appears that these visualizations were either highly ranked or lowly ranked with little in-between, making it difficult for the rankings to be differentiated. In the discussion following their user experiments in data visualization in AR Bach \emph{et al.}~\cite{Bach2018} found that individual difference (of participants) may play a factor in results and these rankings also suggest this.  Looking at the individual user comments and feedback, as well as the result of the VSI questionnaire, helps to provide context.
 
\subsection{User Feedback}
The users who rated the arrow visualization highest stated it provided ``\emph{additional information of direction}'' and that it did not ``\emph{hide details of physical objects}'', and the arrows ``\emph{give you more information''}.
However, those who ranked it lowest stated that there was ``\emph{too much information}'' and their ``\emph{direction was not clear}''.
One user who ranked the banded first and the arrows second stated that the arrows did not make the boundaries clear, but the banded scenarios made it easier to see the limit of the radiation.
Overall, the arrow visualization was the most commented on in the user feedback and had the highest ratio of  positive comments to negative comments (8 positive and 5 negative), as determined positive or negative by the authors. The circle visualization had the most negative comments (9), but surprisingly 3 users had positive comments (``\emph{have a sense of the physical space}", ``\emph{allow me to be more aware of the surroundings}", ``\emph{I like the cases where only some circles are shown, not too bright colors, and not covering too much space}'').
Of the 6 negative comments for the continuous visualization, 4 of them mentioned difficulty in seeing the correct card on the table.
Despite the quantitative results, examining the users' free text responses does indicate some advantages and disadvantages of the various techniques. A further study may be able to reflect the described advantages of the arrow stencil, and better capture the disadvantages of the more solid visualizations.

\subsection{Consideration of Visual Spatial Imagery Ability Questionnaire}

We removed 8 outliers from the data, who ranked themselves as lower than "somewhat agree", as we felt this indicates that they possess lower visual spatial imagery ability. However, this resulted in no significant difference with respect to results in terms of performance.  We also explored the correlation between the average rating a person gave themselves on the VSI scale and their results, using a Kendall correlation due the non-normal nature of the data. Using a range of 0.26 to 0.49 as a definition of moderate (taken from \cite{wicklin2023} which was derived from the Pearson correlation classifications of Schober and Schwarte \cite{Schober2018}),  there is a moderate correlation between the Average of an individual’s VSI score and their average score for the SSM questions across all visualization types (Kendall coefficient of 0.262). 
\begin{table*}
	\caption{Moderate Kendall correlation with the Average VSI score of each participant.}
        \Description{A table showing moderate Kendall correlation with the Average VSI score of each participant for 7 metrics}
	\label{tab:preExp}
		\begin{tabular}{c c c }
		\toprule
		Result Name & Kendall Correlation Coefficient  \\ 
		\midrule
		  Performance Time - Continuous  & 0.373173 \\
		   Average SSM score - Continuous  & 0.329774  \\ 
			Performance Time - Banded 	&  0.305324 \\ 
				Performance Time - Hexagon &  0.305324 \\ 
				Q01 Response - Circle &  0.294504 \\ 
				Q01 Response  - Continuous &  0.265854 \\ 			
				Average SSM score (All visualizations) &  0.261895 \\ 
				&   \\ 
		\bottomrule
		\end{tabular} 
\end{table*}

We considered the results of the SSM questionnaire (average of the answers) and the results of performance in terms of time for each visualization, as well as how the participant answered to Q01, Q02 and Q03, concerning the awareness or the virtual, physical, and balance of both. There were 6 moderate correlations, the strongest being with performance time. This can be interpreted as the higher the persons rating in the VSI pre-questionnaire, the more likely they were to take longer at the trials for the continuous visualization type. They were also to more likely rate it higher on average in the questions for the MEC SPQ. Considering Q01 response, again Continuous had a moderate correlation. The continuous visualization is the one with the most coverage of the mesh and largest set of colors. The banded and hexagon visualizations also had a moderate correlation, and these two visualizations also cover most of the physical world. It appears that those with higher VSI self-ratings are more likely to take longer, particularly for the visualizations that have the most coverage of the spatial awareness mesh. However, the reason behind this is unclear, and further research is required before drawing any firm conclusions.

\subsection{Experiment Hardware and Limitations}
The HoloLens 2 is an OST headset.  However, as devices improve in terms of resolution, processing power, and ergonomic profile, pass-through AR may be more viable option, and additionally it has its own benefits. With OST devices environmental illumination impacts perception and there is no difference between rendered black and full transparency \cite{Erickson2020}.  Using transparency is equivalent to making a color darker and the device has no concept of the real-world color underneath the pixel for alpha blending. Despite the limited effect of the transparency, it still did not make it significantly worse than any of the other visualizations. 
However, a pass-through AR device would more easily allow more complex blending of colors and better use of transparency, using the alpha blending functionality of the graphics hardware. This may be useful with respect to concerns about contrast for the stencil visualizations which were specifically mentioned, by one participant. 
However, recent work by Zhang \emph{et al.}~\cite{Zhang2022} looked at providing improved color contrast between virtual and real objects using an OST device. They capture  background video from the device and blur it to use as a background color to improve color contrast. Such a technique could be explored to help resolve any possible contrast issues with stencils, to improve color blending for transparency, and to improve perceptual consistency of the radiation visualization across a variety of scene backgrounds. Such an approach may have a trade off in performance and would need to be evaluated thoroughly for this use case. Zhang \emph{et al.} do note  that that software-based solutions still have issues overcoming hardware limitations of OST-HMDs, for example, for drawing black or darker colors.
Langlotz \emph{et al.} \cite{Langlotz2016}  suggest an approach for mitigating the effects of undesired color-blending for OST-HMDs, however it depends on custom novel prototype HMD, and is not applicable to commodity HMDs. 

\subsection{Study Design}
This system is visualizing data with the primary aim of influencing the physical movement of users.
This differs from the more traditional visualization techniques where the objective is more concentrated on purely understanding the data. Here the data must first be understood, then the user should move around accordingly. In this study there was no interaction with the underlying data. These factors perhaps shaped the user behavior and hence the results.
 
The results were often not significant between each condition. However, we do note that one of our own visualization questions provided a significant difference, even if it was not part of a hypotheses. The visual spatial imagery questionnaire provided some moderate correlations, thus pointing to it having some relevance. Other results were largely non-significant, but the approaches were chosen as they are already widely used in the mixed reality community or in the wider domain of situation awareness. Given these points, we feel there is a specific need for a situated visualization questionnaire which can more accurately explore the impact of  the visualization not only on understanding the data, but also the relationship between levels of situation awareness, and the ultimate effects on human movement behavior.  

While the experimental conditions changed (visualizations) the room layout did not. Therefore, it is not possible to say if the room layout itself was a factor in the largely non-significant results. 
Furthermore, the number of test conditions was relatively large, combined with the fact that except for the card task, the scenario was largely static. All these combined with a simple task, and an identical room layout probably influenced user behavior. These aspects should be examined in more detail through further studies.

The motivation of the card tasks was to force the user to search for a real-world item, showing they could perceive both the real and the virtual. However, the 52 degree AR display area of the HoloLens 2 means that a user may be able to view the real-world elements (such as the card table) without looking  through the AR portion of the display. Participants were instructed to only look for the card through the AR display, however it was not possible to strictly enforce this, as it was not possible for the experimenter to be in the room close to the participant. One participant even made the comment ``\emph{the headset was a bit small and the face of the cards could be seen directly by looking down}''.

In our experiment a single source had a radius of 2 meters. This was to avoid having visualizations completely dominate the experiment room. Looking at the mean nearest source distance in Figure \ref{fig:result_mean_dist} it can be seen that users in general kept quite far, with the median mean distances being in the range of 1.8 to 1.9 meters. It may be possible that participants were using the edges of the visualization as a barrier to guide them. One participant did comment that the arrow visualization made it ``\emph{difficult to find the boundary}'' and that banded visualization allowed then ``\emph{to better see the limit of the radiation}'', which suggests that the visualization boundary was an important feature for them. Adding clearly defined boundaries is a possible future design feature for the arrow based visualizations. 

\subsection{Recommendations and Lessons Learned}
Based on our experiences of the experiment, we make recommendations in the areas of visual encoding, study design, and individual differences.
\subsubsection{Visual Encoding}
In terms of visual encoding more information does not necessarily result in better performance.  The inclusion of arrows explicitly specifying direction did not improve performance significantly. It also had no negative impact either. Therefore, it may be included as part of a visualization, but the context of the visualization and the preferences of the specific users of a system should be considered and balanced against the cost of implementation and processing time.  
\subsubsection{Study Design}
The study was carefully designed to ensure that participants used information from the visualizations to navigate. However, visualization that affects behavior may need a different approach to visualization that imparts knowledge. Our participants typically completed the tasks very quickly, limiting their exposure time.  While this is desirable, it remains to be seen if the same duration would arise when experts use the system. Furthermore, the short times resulted in a small effect size for time, and therefore radiation dose, making statistical difference more difficult to achieve as part of an evaluation. We recommend exploring tasks where the desired behavior modification can be identified more explicitly. 
It may be worth examining if an effect size can be detected when undertaking a radiation source type identification scenario, in the context of CBRN response.

Additionally, the high number of trials conducted means that participants may have desired from a purely personal and non-CBRN level to avoid boredom and complete the task as quickly as possible. In future, we may explore fewer variations and iterations of the trial for each user, plus explore a longer task. However, this is quite challenging, as adding more trials or lengthening trial duration to improve the effect size may have other negative impacts on results.
The study did not contain a no visualization option, this was a design choice. However, it means that we are not able to assess the effectiveness of visualization in general.

Choice of questions plays an important role, asking the wrong questions may obscure an interesting result.
Due to our focus on situation awareness we decided to adapt questions from established questionnaires in the literature. In retrospect our choices should have been more focused on our specific experiment. In Figure \ref{fig:questionnaire_results}, it can be seen that, generally, the custom questions for this experiment have a wider spread of results, and that the MEC and SART questions were less effective at distinguishing the visualizations. Additionally, the pre-questionnaire was focused on VSI ability, however other questions could have been included, for example relating to experience with video games. One participant mentioned they found the hexagons less intrusive, but this may have been because of ``\emph{some hexagon based games}'' they had played.
\subsubsection{Individual differences need more exploration.}
This study did not go into detail on individual differences, either through conducting spatial ability tests, or analyzing multiple groups based on the MEC VSI questionnaire. However, the ranking data points to their being strong differences in preferences, with some binary distributions occurring. Therefore, a more thorough study which explores the role of individual differences is required.
The comments from users indicated some designs and levels of transparency may have had an impact on task behavior. While this was not noted by all users, it points to the need perhaps to change certain features or perhaps allow customization depending on user preference. For example, it is not clear if all users (there may be many) require the same visualization technique. Instead, they could choose one which they find the most effective, and/or perhaps have customization options e.g., arrow size, transparency etc.
Some user comments, such as continuous visualization obscuring the card table, were not surprising. However, other aspects such as the bimodal distribution of preference rankings for some visualizations, and the comment about the arrows not making radiation boundaries clear were surprising and will help steer future work.

\section{Conclusions}
At the technical level, we have described an approach that allows users in AR to visualize data in the physical environment around them leveraging the spatial awareness mesh. We have demonstrated that GPU based shading of the spatial awareness mesh a technically viable approach to visualizing data in real time in a training scenario. Our fundamental approach may be applied to other use cases beyond our scenarios for radiological incident response.

We designed and performed a user experiment to compare visual encodings with the aim of understanding which approach could help the user maintaining awareness of both the virtual threat and the physical world in a balanced fashion. Our results show that enhancements to the fundamental approach of coloring the spatial awareness mesh to improve user performance and balance perception of the virtual and physical environment do not make a significant difference. However, visualizing data in a spatial environment is challenging and more research is required to understand the confounding aspects that might be at play. Individual differences, between participant in experiments may need to be better understood. Finally, we have a detailed discussion around evaluation approaches for in-situ visualization.

\bibliographystyle{ACM-Reference-Format}
\bibliography{AR_Radiation_Visualization}

\end{document}